\documentclass[twoside,twocolumn,9pt]{article}
\pdfoutput=1
\usepackage{extsizes}
\usepackage[super,sort&compress,comma]{natbib} 
\usepackage[version=3]{mhchem}
\usepackage[left=1.5cm, right=1.5cm, top=1.785cm, bottom=2.0cm]{geometry}
\usepackage{balance}
\usepackage{mathptmx}
\usepackage{sectsty}
\usepackage{graphicx} 
\usepackage{lastpage}
\usepackage[format=plain,justification=justified,singlelinecheck=false,font={stretch=1.125,small,sf},labelfont=bf,labelsep=space]{caption}
\usepackage{float}
\usepackage{fancyhdr}
\usepackage{fnpos}
\usepackage[english]{babel}
\addto{\captionsenglish}{%
  
}
\usepackage{array}
\usepackage{droidsans}
\usepackage{charter}
\usepackage[T1]{fontenc}
\usepackage[usenames,dvipsnames]{xcolor}
\usepackage{setspace}
\usepackage[compact]{titlesec}
\usepackage{extsizes}
\usepackage[super,sort&compress,comma]{natbib} 
\usepackage[version=3]{mhchem}
\usepackage[left=1.5cm, right=1.5cm, top=1.785cm, bottom=2.0cm]{geometry}
\usepackage{balance}
\usepackage{longtable}
\usepackage{seqsplit}
\usepackage{times,mathptmx}
\usepackage[hidelinks]{hyperref}
\usepackage{subcaption}
\usepackage{booktabs}
\usepackage{makecell}
\usepackage[format=plain,justification=raggedright,singlelinecheck=false,font={stretch=1.125,small,sf},labelfont=bf,labelsep=space]{caption}
\usepackage[usenames,dvipsnames]{xcolor}
\usepackage{setspace}
\usepackage{multirow}
\usepackage[compact]{titlesec}
\usepackage{wrapfig}
\usepackage{tabularx}
\usepackage{multirow}

\usepackage{epstopdf}

\definecolor{cream}{RGB}{222,217,201}

\begin{document}

\pagestyle{fancy}
\thispagestyle{plain}
\fancypagestyle{plain}{
\renewcommand{\headrulewidth}{0pt}
}

\makeFNbottom
\makeatletter
\renewcommand\LARGE{\@setfontsize\LARGE{15pt}{17}}
\renewcommand\Large{\@setfontsize\Large{12pt}{14}}
\renewcommand\large{\@setfontsize\large{10pt}{12}}
\renewcommand\footnotesize{\@setfontsize\footnotesize{7pt}{10}}
\makeatother

\renewcommand{\thefootnote}{\fnsymbol{footnote}}
\renewcommand\footnoterule{\vspace*{1pt}%
\color{cream}\hrule width 3.5in height 0.4pt \color{black}\vspace*{5pt}} 
\setcounter{secnumdepth}{5}

\makeatletter 
\renewcommand\@biblabel[1]{#1}            
\renewcommand\@makefntext[1]%
{\noindent\makebox[0pt][r]{\@thefnmark\,}#1}
\makeatother 
\renewcommand{\figurename}{\small{Fig.}~}
\sectionfont{\sffamily\Large}
\subsectionfont{\normalsize}
\subsubsectionfont{\bf}
\setstretch{1.125} 
\setlength{\skip\footins}{0.8cm}
\setlength{\footnotesep}{0.25cm}
\setlength{\jot}{10pt}
\titlespacing*{\section}{0pt}{4pt}{4pt}
\titlespacing*{\subsection}{0pt}{15pt}{1pt}

\fancyfoot{}
\fancyfoot[LO,RE]{\vspace{-7.1pt}\includegraphics[height=9pt]{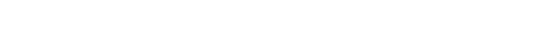}}
\fancyfoot[CO]{\vspace{-7.1pt}\hspace{13.2cm}\includegraphics{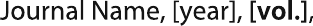}}
\fancyfoot[CE]{\vspace{-7.2pt}\hspace{-14.2cm}\includegraphics{head_foot/RF}}
\fancyfoot[RO]{\footnotesize{\sffamily{1--\pageref{LastPage} ~\textbar  \hspace{2pt}\thepage}}}
\fancyfoot[LE]{\footnotesize{\sffamily{\thepage~\textbar\hspace{3.45cm} 1--\pageref{LastPage}}}}
\fancyhead{}
\renewcommand{\headrulewidth}{0pt} 
\renewcommand{\footrulewidth}{0pt}
\setlength{\arrayrulewidth}{1pt}
\setlength{\columnsep}{6.5mm}
\setlength\bibsep{1pt}

\makeatletter 
\newlength{\figrulesep} 
\setlength{\figrulesep}{0.5\textfloatsep} 

\newcommand{\topfigrule}{\vspace*{-1pt}%
\noindent{\color{cream}\rule[-\figrulesep]{\columnwidth}{1.5pt}} }

\newcommand{\botfigrule}{\vspace*{-2pt}%
\noindent{\color{cream}\rule[\figrulesep]{\columnwidth}{1.5pt}} }

\newcommand{\dblfigrule}{\vspace*{-1pt}%
\noindent{\color{cream}\rule[-\figrulesep]{\textwidth}{1.5pt}} }

\makeatother

\twocolumn[
  \begin{@twocolumnfalse}

\vspace{1em}
\sffamily
\begin{tabular}{p{18cm} }

\noindent\LARGE{\textbf{Electrolyte Coatings for High Adhesion Interfaces in Solid-state Batteries from First Principles}}
\\
\vspace{0.3cm} \\

 \noindent\large{Brandi Ransom\textit{$^{a}$}, Akash Ramdas\textit{$^{a}$}, Eder Lomeli\textit{$^{a}$},  Jad Fidawi\textit{$^{a}$}, Austin Sendek\textit{$^{a,c}$}, Thomas Devereaux\textit{$^{d}$}, Evan Reed\textit{$^{a}$}}, and Peter Schindler\textit{$^{b}$}. \\

 \\
 \noindent\normalsize{We introduce an adhesion parameter that enables rapid screening for materials interfaces with high adhesion. This parameter is obtained by density functional theory calculations of individual single-material slabs rather than slabs consisting of combinations of two materials, eliminating the need to calculate all configurations of a prohibitively vast space of possible interface configurations. Cleavage energy calculations are used as an upper bound for electrolyte and coating energies and implemented in an adapted contact angle equation to derive the adhesion parameter. In addition to good adhesion, we impose further constraints in electrochemical stability window, abundance, bulk reactivity, and stability to screen for coating materials for next-generation solid-state batteries. Good adhesion is critical in combating delamination and resistance to Lithium diffusivity in solid-state batteries. Here, we identify several promising coating candidates for the $Li_7La_3Zr_2O_{12}$ and sulfide electrolyte systems including the previously investigated electrode coating materials $LiAlSiO_4$ and $Li_5AlO_8$, making them especially attractive for experimental optimization and commercialization.} \\

\end{tabular}

 \end{@twocolumnfalse} \vspace{0.6cm}]


\renewcommand*\rmdefault{bch}\normalfont\upshape
\rmfamily
\section*{}
\vspace{-1cm}

\footnotetext{\textit{$^{a}$~Department of Materials Science and Engineering, Stanford University}}
\footnotetext{\textit{$^{b}$~Department of Mechanical and Industrial Engineering, Northeastern University}}
\footnotetext{\textit{$^{c}$~Aionics, Inc.}}
\footnotetext{\textit{$^{d}$~Stanford Institute for Materials and Energy Sciences, Stanford University}}

\section{Introduction}
The battery market is experiencing incredible growth with projections showing no sign of slowing down.\cite{batterymarket} Electric vehicles, mobile electronics, grid-scale renewable energy farms, implantable medical devices, and more are incredibly diverse examples of devices relying on batteries. As electrification becomes the norm, these various utilities of batteries will require different performance optimization of cycle life, operating voltage, and power. In many previous works battery materials have been investigated for maximizing these properties among others, and the methods developed therein could be repurposed to target specific values, not just the maxima.\cite{matdisc1,matdisc2,matdisc3} The vast growth of the battery market would also be easier to sustain with a more diverse set of battery materials. We have seen the detrimental environmental and social effects from the dependence on Cobalt for Li-Co-O cathodes.\cite{shapshak_2019} A few solid-state materials optimizations have begun to take this into account when presenting novel materials for battery application, showing the viable possibilities of high performance solid-state materials with comparable performance to liquid batteries.\cite{INDU2021100804} These novel materials searches are pushing the solid-state field forward in optimization techniques and our understanding of structure-electrochemical performance relationships, however novel materials alone are not enough to upend the liquid battery industry. 

Solid-state battery success will also likely depend on a Li-metal anode to become commercially viable in their energy density performance. Due to Lithium metal’s high reactivity, there are very few materials which can come into contact with Li metal without reacting, typically Lithium binary salts, which do not have a high enough ionic conductivity.\cite{xiao_wang_bo_kim_miara_ceder_2019} However, ionic conductivity and thermodynamic stability are bulk properties, which are insufficient to predict the performance of an operational battery cell because of the critical impacts of interfacial interactions on performance. The key to success in liquid batteries is the self passivating solid electrolyte interphase (SEI) layer between the electrolyte and the electrodes, a concept which has only begun to be investigated in the solid-state field. Further development towards finding solid materials that are not reactive at the interface is critical.\cite{CAMACHOFORERO2018782} While reactivity is a necessary constraint for any battery, it does not account for the operational voltage or interfacial resistance requirements that are also necessary. Coatings are highlighted as the solution to meeting both requirements, allowing a much broader set of materials to be combined for successful batteries in situ. 

Computationally generated thermodynamic data has been readily used for electrochemical compatibility between solid electrolytes and electrodes, a necessary first step for describing the constraints on chemistries of an all solid-state battery.\cite{zhu_he_mo_2016} These constraints highlight the physical gaps in electrochemical stability between lithium metal anodes and high voltage cathodes that can only be filled by coatings with electrochemical stability windows large enough to overlap the full operational range of the electrode. However, two materials that are chemically non-reactive and are stable within the same voltage ranges can face large interfacial resistances.\cite{kim_rupp_2020} Solutions to interfacial resistance involve maximizing interfacial contact between materials, which has been explored on the macroscale with various synthesis and assembly techniques.\cite{jiang_han_wang_wang_2019} 

Strong adhesion is thought to lead to longer mechanical life in batteries, as it helps maintain a well contacted interface throughout cycling and volume changes of battery components. This was shown experimentally by applying pressure to cells, forcing interfaces to be better wetted, or adhered.\cite{schlenker} In solid-state batteries, increasing the area of contact between materials is critical to taking advantage of the bulk ionic conductivity of promising Li-P-S and Li-La-Zr-O systems. These systems are also promising in that the Li-P-S system has previously been investigated for high interfacial contact due to it's low Young's modulus, and $Li_7La_3Zr_2O_{12}$ has shown the most promising stability in the field.\cite{zhou_zhang_shen_fang_kong_feng_xie_wang_hu_wang_etal._2022,fan_ji_han_yue_chen_chen_deng_jiang_wang_2018,connell_fuchs_hartmann_krauskopf_zhu_sann_garcia-mendez_sakamoto_tepavcevic_janek_etal._2020} However, to realize these compounds to have the same 4 Volt electrochemical stability window we expect from liquid electrolytes, we would need solid coatings between these electrolytes and their electrodes. 

Adhesion strength has been experimentally characterized by contact angle measurements of a drop of liquid on a solid surface\cite{contact_angle} or by various mechanical tests (stud pull test, blister test, scribe method, tape-peel test, among many others). However, it is challenging to extract an accurate, quantitative measure of adhesion strength from these methods. Instead, for thin-film coatings a laser-induced spallation technique has shown to be a more quantitative non-contact adhesion measurement method.\cite{pete1}  There has also been recent advances in first principles workflows to determine the adhesion strength by calculating the potential energy surface of interfaces using density functional theory.\cite{WOLLOCH2022111302} However, this has only been done for elemental crystals\cite{pete2} (and a few select binary crystals) as the space of possible interface configurations is prohibitively vast for interfaces of compounds with more than one chemical constituent.

This work focuses on the development of an effective approximation of adhesion between solid materials, and uses the screening of coatings for solid-state batteries as the test canvas due to the breadth of materials relevant to the space and the rapid progression of the technology. Specifically, we explore coatings that would be stable with the ideal Li-metal anode, well researched cathodes, and the promising electrolyte systems (LLXO, X = Zr or Ta), Li-P-S (LPS), and Li-B-S (LBS) (defined in section \ref{sec:DD}). We prioritize common materials in our screening to show the chemically diverse and viable options for all solid-state battery chemistries. The approximation allows for extremely fast narrowing of the viable candidate space for any problem in which adhesion would benefit the end goal. The continuous nature of the adhesion parameter allows for ranking of all materials of interest for direct comparison. High-adhesion coatings with interfacial compatibility provide the physical bridge to realize higher voltage cathodes, metal anodes and better performing interfaces for next-generation solid-state batteries.

The remainder of this paper proceeds as follows. Sections 2 and 3 establish the assumptions and boundaries of our approximation of the calculated adhesion parameter, outlines the data used in our screening process, and describes each of the solid electrolyte systems chosen. Section 4 establishes the screening progression and overall viability of coatings for each of the solid electrolyte systems.

\section{Materials and Methods}
The methods developed throughout this work were applied specifically to solid-state interfaces of battery-compliant materials. This is only an application of the adhesion parameter, though the methods can be applied to other solid-solid interface use cases. 
\subsection{Data Description}\label{sec:DD}

Our pool of data originates from the Materials Project database (Extracted: March 2022), which contains structures and Density Functional Theory (DFT)-based data regarding 19480 lithium containing compounds.\cite{jain_ong_hautier_chen} These materials were the starting candidates for coatings which could be applied to the cathode or anode side of a solid electrolyte. Three specific chemical systems (LBS,LPS,LLXO) were chosen as sample solid electrolytes for this work. These were chosen because of the increased focus on optimization of these compounds for commercialization in the literature, to show that the materials selected as “best candidates” in this work may be worth further pursuit. The systems, specific electrolytes, and their materials project ID numbers are listed here.  
\begin{table}[H]
\centering
\begin{tabular}{ll}
        $Li_5B_7S_{13}$, mp-532413&  $Li_{10}Ge(PS_6)_2$, mp-696128\\
         \textbf{$Li_2B_2S_5$, mp-29410} & $Li_3PS_4$, mp-1097036\\
         \textbf{$Li_3BS_3$, mp-5614}&\textbf{$Li_7P_3S_{11}$,  mp-641703}\\
         \textbf{$Li_7La_3Zr_2O_{12}$, mp-942733}  & \textbf{$Li_5La_3Ta_2O_{12}$, mp-559776} \\

    \end{tabular}
\end{table}
Materials in bold have experimentally validated structures. Materials which haven't been experimentally verified by the Materials Project database were matched to literature based on space group. 

\subsection{Screening Metrics for Coating-Electrolyte Compatibility} \label{screening}
While this work proposes a parameter that can be used to approximate the adhesion between any two solid materials, we additionally impose other basic constraints on candidate coating materials in order to provide analysis on plausible materials compatible in solid-state batteries. We choose the following constraints as necessary theoretical criteria for an operational cell. All criteria can be found or calculated using functionalities of the materials project API and pymatgen.\cite{MP_api,PhysRevB.84.045115} 

\begin{figure}[h]
    \centering
    \includegraphics[width=0.45\textwidth]{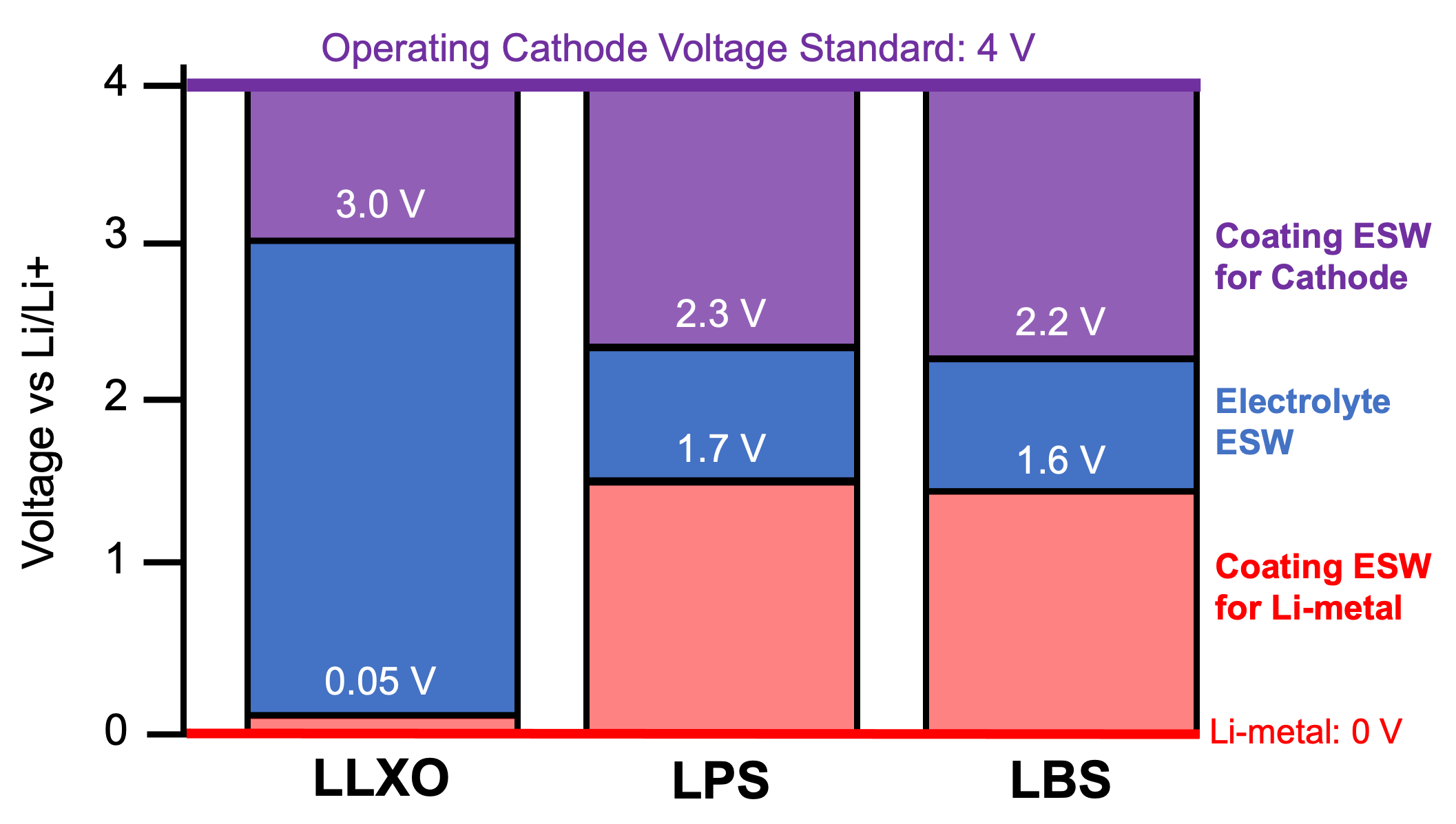}
    \caption{Schematic of the required electrochemical stability windows of coatings for each of the electrolyte families of interest.}
    \label{fig:ESW_schematic}
\end{figure}

As previously described, we first only consider materials which contain lithium. This is a necessary criterion for coating materials as an extension of the electrolyte functionality to prevent a decrease in capacity of the battery from lithium absorption by the coating material. As an extension of the electrolyte\footnote[1]{Placement of the coating within the solid-state battery briefly discussed in the appendix.} we ensure materials have a band gap larger than 2 eV, to prevent electronic conductivity, which would expose the electrolyte to the potential that it is supposed to be shielding.  In hopes of easing the experimental exploration of any candidates discussed in the work, we also screen out materials with elements above atomic number 80, Lanthanides. Heavy materials can severely decrease the gravimetric energy density, and also tend to be less abundant, therefore the coatings should be as light as possible. We further only consider candidates which lie on the convex hull of their chemical systems, or $E_{hull}=0$, implying that the candidates will not decompose on their own accord. This work focuses on applicability of the adhesion parameter for approximation, and Li-metal/ceramic systems are chosen as an exciting application space.
For coating candidates not to hinder the operation of a cell, the electrochemical stability window (ESW) of the coating must fully span the ESW of the electrode and overlap with the window of the electrolyte. Using $Li/Li+$ as the reference material, the Li-metal electrode has an operational ESW of $0~ V$, and we take a cathode standard to reach an operational voltage of $4~V$. This cathode standard encompasses a range of top voltages of commonly employed cathode materials ($FePO_4 = 3.4~V$, $LiCoO_2 = 4.2~V$, $LiMnO_2 = 4.1~V$,$LiNiO_2 = 4.1~V$).\cite{zaghib_dub_dallaire_galoustov_guerfi_ramanathan_benmayza_prakash_mauger_julien_etal._2012,kalluri_yoon_jo_park_myeong_kim_dou_guo_cho_2017,ding_xie_cao_zhu_yu_zhao_2010,park_zhu_torres} It has been readily investigated that thermodynamic ESWs are underestimations of stability, allowing our criteria to act as a strategic lowerbound to ensure that the chosen candidates will be stable insitu.\cite{schwietert_vasileiadis_wagemaker_2021} The ESWs of each candidate are computed by constructing the convex hull of the grand potential phase diagram as a function of applied Li chemical potential using methods of Ong et al. and the phase\_diagram module of pymatgen.\cite{ong_wang_kang_ceder_2008} Figure \ref{fig:ESW_schematic} shows the ESW requirements of coatings for the three electrolyte systems we have chosen based on experimental values discussed in these references for each family: LLXO\cite{han_zhu_he_mo_wang_2016}, LBS\cite{sendek_antoniuk_cubuk_ransom_francisco_buettner-garrett_cui_reed_2020}, LPS\cite{schwietert_vasileiadis_wagemaker_2021}. 

In addition to individual materials' properties, we consider the most basic requirements of two materials at the interface, chemical interfacial stability. The reaction energy between each candidate and its respective electrolyte or electrode are computed by the interface\_reactions module in pymatgen.\cite{pymatgen} This module fits the stoichiometries of possible reactions between a material system, and outputs the reaction energy calculated by the weighted formation energies of the materials. We take the thermodynamic driving force of each pair of materials to come from the stoichiometric combination of the materials which maximizes the exothermic energy release. This value acts as an upper bound on the driving force of bulk materials to react when placed within the same system. We screen for materials with a maximum driving force of greater than -0.1 eV between the coating and electrolyte, accounting for kinetic stability\cite{PhysRevB.84.045115}, again only presenting the materials that have the highest viability for experimental success. 

Three further properties which are integral to battery operation and production are ionic conductivity, elemental cost, and elemental abundance. Ionic conductivity in electrolyte materials was the paramount barrier for replacing liquid with solid electrolytes and we incorporate experimentally validated ionic conductivities into our analysis of the identified well-adhering compounds. Elemental cost and elemental abundance have both economic and social effects as  seen most notably with Cobalt.\cite{shapshak_2019} As the primary goal of this paper is to present a successful approximation for adhesion between materials, we discuss coatings for each electrolyte based on their performance of adhesion, particularly analyzing any trends in anionic groups of the high-performing coatings. In featuring the full list of viable coatings, many other considerations, including the three discussed here, can be taken into account when selecting which materials to pursue further.

\subsection{Construction of the Adhesion Parameter} \label{ad_param}
Historically, adhesion is characterized by the contact angle of a drop of liquid on a solid surface,\cite{contact_angle} with the analogous solid-solid interface described by the same relationship.\cite{WINTERBOTTOM1967303} The larger the contact angle, the worse the adhesion between the two materials. In solid systems, this contact angle is commonly converted into a relation of surface energies between the two materials in the following expression: 
\begin{equation}
    \gamma_{e} = \gamma_{ec}+\gamma_{c}\cos{\theta}.
    \label{eq:CASE}
\end{equation} 
Where $\gamma_{e}$ and $\gamma_{c}$ represent the electrolyte and coating surface energies, respectively, and $\gamma_{ec}$ represents the electrolyte-coating interfacial energy for our system. We approximate "good" adhesion as $\theta\leq90~\deg$, and a perfectly adhered system would have $\theta=0$. Within any bulk material, the surface energy could be described by any of the various terminations and Miller orientations that can be created through a unit cell. In our work we calculate the surface energy of all possible terminations and orientations by DFT. Details of these calculations can be found in the appendix.

Surface energy values have a lower bound of 0, as it always takes energy to break bonds to create the surface. However interfacial energies can be positive or negative, and the lower (more negative) values represent more stable, and therefore perhaps more likely naturally occurring interfaces. We hypothesize that more stable surfaces are less likely to bond or adhere to other materials. In the derivation below $min(\gamma_{x})$ refers to the most stable termination among all terminations and orientations (up to Miller index 1) of material $x$. We utilize $min(\gamma_{x})$ and $max(\gamma_{x})$ to establish our adhesion parameter approximation as a lower bound on adhesion between two solid materials.  

We begin by isolating $\cos{\theta}$, and stating its bounds:

\begin{equation}
 \frac{\gamma_{e}-\gamma_{ec}}{\gamma_{c}}=\cos{\theta},
    \label{eq:CASE2}
\end{equation}
where $-1\leq\cos{\theta}\leq1$. The true value of the ratio of surface energies is limited by bounded cases of each individual surface energy. These bounds exist because of the various possible terminations or exposed surfaces of each material.
\begin{equation}
 \frac{\gamma_{e}-\gamma_{ec}}{\gamma_{c}}\leq\frac{max(\gamma_{e})-min(\gamma_{ec})}{min(\gamma_{c})}.
    \label{eq:CASE3}
\end{equation}

We substitute $\cos{\theta}$ from equation \ref{eq:CASE2} into equation \ref{eq:CASE3} and isolate $\gamma_{ec}$ as the variable of interest. 
\begin{equation}
min(\gamma_{ec}) \leq max(\gamma_{e})-min(\gamma_{c})\cos{\theta}
    \label{eq:CASE4}
\end{equation}

In order to achieve the most favorable interfacial interaction, represented by the lowest $\gamma_{ec}$, we aim to minimize the right side of the equation which will act as an upper bound on $min(\gamma_{ec})$. To find materials which are best adhered, we use the case of perfect adhesion between two materials $\cos{\theta}=0$. For adhesion to be favorable, $\gamma_{ec}<0$, and therefore the goal of this work is to identify materials for which
\begin{equation}
max(\gamma_{e})-min(\gamma_{c})\leq 0.
    \label{eq:CASE5}
\end{equation}
The left side of equation \ref{eq:CASE5} will be further referred to as the "adhesion parameter" in units of $eV/$\AA$^2$. This provides an approximation for adhesion from values which can be quickly calculated from quantum mechanical methods for solid materials. A detailed description for our approximation of $\gamma$ is in the appendix.

\begin{figure*}[h]
    \centering
    \includegraphics[width=0.6\textwidth]{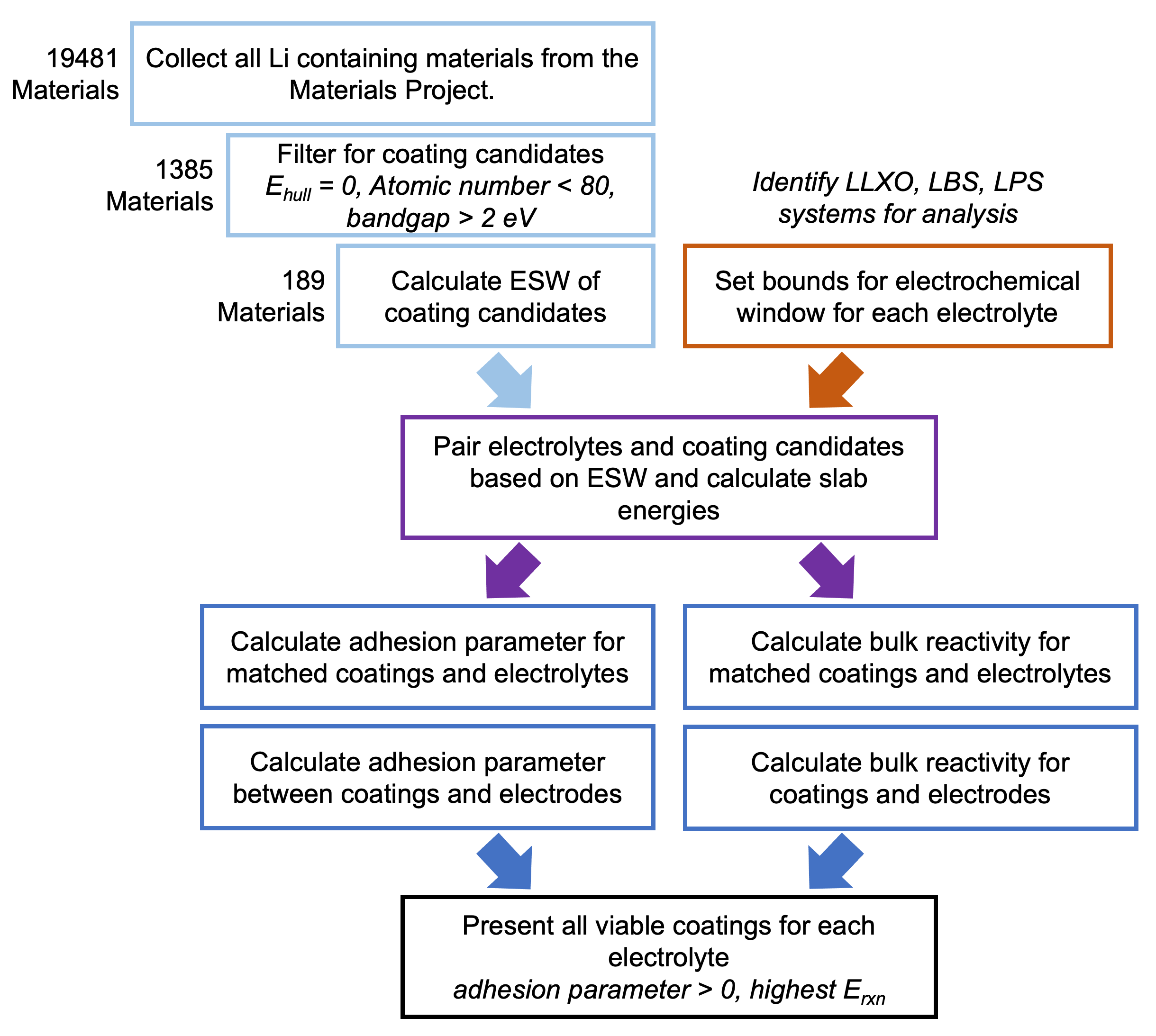}
    \caption{This graphic outlines the order of filters applied to and calculations done on our coating candidate data set.}
    \label{fig:screeningprocess}
\end{figure*}

\section{Results}
Figure \ref{fig:screeningprocess} describes the process of data collection, processing, and analysis for this work. The order in which candidate materials were filtered out was chosen to minimize computational expense. 
Then the adhesion parameter is calculated on all materials which pass these necessary markers for baseline battery operation, for further analysis and direction on the types of materials that could have higher performance in batteries. Additionally, We include the bulk reactivity in our analysis, as described in section \ref{screening}. The best materials across all electrolytes are discussed in section \ref{sec:discussion} of this paper, and we also provide the entire list of candidate coatings for which we calculate a favorable adhesion parameter with each electrolyte in the Appendix. 
\subsection{Materials Screening }
Beginning from 19481 Lithium containing materials in the Materials Project, the atomic number and $E_{hull}$ filters quickly reduce our candidate list to 1385 candidates. After pairing the ESWs, the electrolyte systems have LLXO:36, LBS:11, and LPS:11 for Lithium metal side candidates, and LLXO:156, LBS:82, and LPS:93 for cathode side candidates. These results contain 156 unique materials, totalling in 945 slab terminations. 15 of these materials were not converged from DFT calculations after $\sim$2 weeks, likely due to extremely unstable or large surfaces of particular slabs. Our analysis of materials is generalized and these few particular materials will not skew our general analysis or prevent us from validating our adhesion parameter approximation, and there are other similar materials in the data set from which we can gain insight.\footnote{These are the materials for which DFT calculations did not converge:$Li_3GaF_6$, $K_2Li_3B(P_2O_7)_2$, $Rb_2Li_3B(P_2O_7)_2$, $CsLi(PO_3)_2$, $Li_2Ge(S_2O_7)_3$, $KLi_3Ca_7Ti_2Si_{12}(O_{18}F)_2$, $Na_3Li_3Sc_2F_{12}$, $LiVZnO_4$, $Li_2CrO_4$, $Li_5TiN_3$, $KLiMoO_4$, $LiZnAsO_4$, $Li_7La_3Hf_2O_{12}$, $LiZnPO_4$,$Li_2MoO_4$}

\subsection{Adhesion Parameter Approximation}
\begin{figure*}[h]
\centering
    \includegraphics[width=0.8\textwidth]{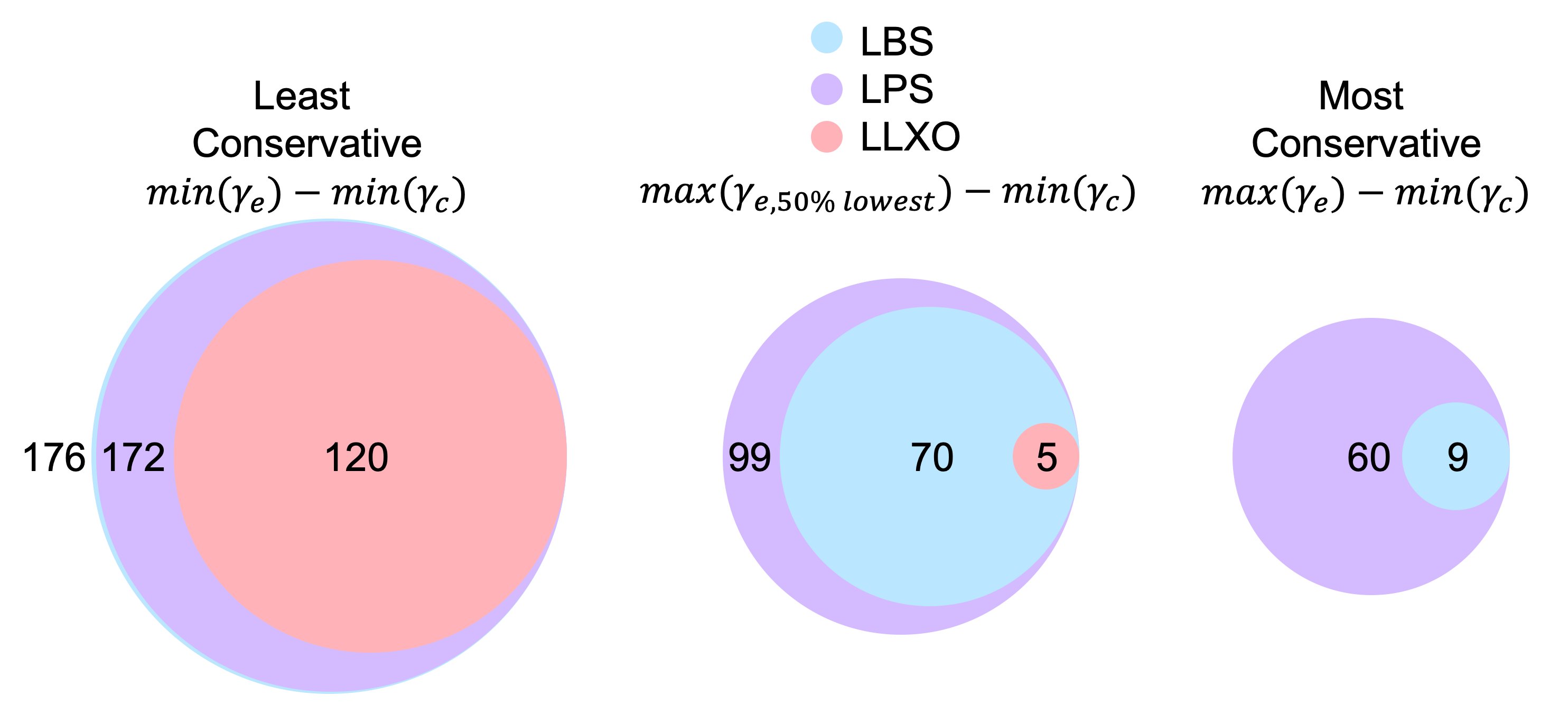}
    \caption{These diagrams show the number of coatings which meet the ESW criteria and have a favorable adhesion parameter with the electrolyte system, for three different levels of conservativeness.}
    \label{fig:venn}
\end{figure*}

The slab with the lowest surface energy for each coating was chosen to represent $\gamma_{c}$ in our adhesion parameter approximation. Our calculation of the adhesion between interfaces is conservative, allowing us to narrow in on materials which have the highest chance of wetting to an electrolyte or electrode. The $max(\gamma_{e})$ value represents the most unstable termination of the electrolyte material, which is unlikely to occur as an exposed surface, both because electrolytes take a polycrystalline form and more stable terminations are more naturally occurring. For this reason we also analyze two less conservative scenarios. First, we instead use the maximum value of the lowest 50\% of surface energies (i.e. the upper bound of the more likely occurring surfaces). Second, we use the most stable termination for all electrolytes, which is the least conservative approximation. The differences in number of materials which align with each of the bounds for $\gamma_{e}$ are shown in figure \ref{fig:venn}. 

This analysis shows a precision-recall trade off for finding well-adhering interfaces. Our most conservative case represents the highest precision, where we have the highest confidence that the interfaces will adhere. Our least conservative case represents the highest recall, ensuring that we have captured all coatings which may adhere, and only screening out those we are most confident will not adhere. From figure \ref{fig:venn} we see that the moderate case best fits our ambition to present promising materials, giving a reasonable number of materials that would allow experimentalists to select which materials to further investigate. In order to account for more stable surfaces being exposed in nature, we proceed with the moderate case, where the 50\% most stable terminations for each material are considered. 

\begin{figure}[h]
    \centering
    \includegraphics[height=8cm]{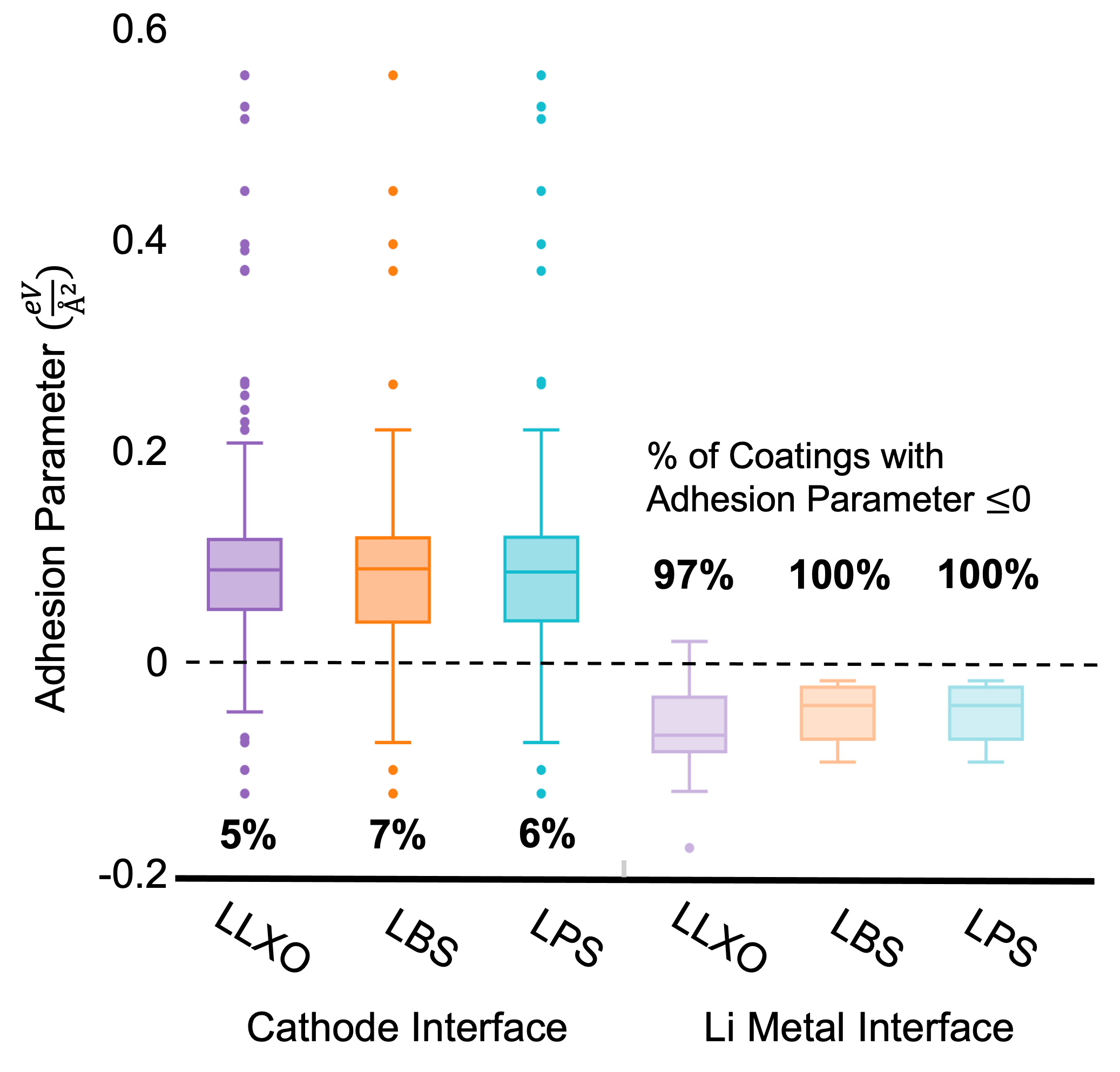}
    \caption{Each box and whisker plot represents the distribution of adhesion parameters at the electrode-coating interface with compatible electrochemical stability windows and meet stability and band gap constraints for the electrolyte system. This includes coatings which have poor adhesion between the electrolyte and coatings. As defined in section \ref{ad_param}, an adhesion parameter less than zero is considered good adhesion. }
    \label{fig:lyteadhesion}
\end{figure}

The results of our adhesion parameter calculations for the electrode-coating interface are shown in figure \ref{fig:lyteadhesion}. We easily identify the disparity in the number of available coatings which would adhere well to the respective electrode. These distributions demonstrate that the limiting interface with adhesion is between the electrolyte and coating for anode coatings, but between the cathode and coating for cathode coatings.  When further considering the electrode-coating adhesion for our final list, the LLXO electrolytes had significantly less well-adhering, non-reactive candidates (5), compared to LBS (33) and LPS (49) systems. The range for means across all electrolytes is very narrow, and the LLXO electrolytes show the lowest (best) adhesion parameter overall. Both $Li_{10}Ge_2S_{12}$ and $Li_5B_7S_{13}$ have significantly fewer promising candidates as other electrolytes in their groups, highlighting that small elemental and structural differences can greatly affect the bonding between surfaces. The fact that we identified more than 70 materials that adhere well to sulfide electrolytes shows the importance of this work, in being able to extend the ESWs and allow for more feasible electrode-electrolyte combinations in batteries. Because we are taking a conservative approach with this approximation, it is highly likely that these distributions underestimate adhesion between materials. 

\subsection{Bulk Reactivity and Electrode Analysis}
As an added metric for stability between candidate coatings and electrolytes or electrodes, we calculate the bulk reactivity ($E_{rxn}$) between pairs of materials. By setting a threshold for $E_{rxn}$ in combination with a high adhesion parameter, we identify the most promising materials which should bond at the interface, but not react to consume each other and form byproducts. From the set of well-adhering candidates, the cutoff of $-0.1~eV$ is limiting for some electrolytes. LLTO has no candidates with $E_{rxn} \geq -0.25~eV$, and other electrolytes have only a few candidate materials that we present for discussion. We then repeat adhesion parameter and $E_{rxn}$ calculations on coatings with either Lithium metal or common cathode materials $FePO_4$, $LiCoO_2$, $LiMnO_2$, $LiNiO_2$. There are very few coatings which can withstand all of these tests to be considered a promising candidate. For our analysis of top materials we focus on materials which were found to be most promising for each electrolyte group. In the Appendix we present all materials with a negative adhesion parameter along with their $E_{rxn}$ values and data for interfacing with cathodes and Lithium metal.

\section{Discussion}\label{sec:discussion}

All materials for which coating candidates had a negative adhesion parameter and an $E_{rxn} > -0.1~eV$ with both an electrolyte and at least one electrode is shown in the appendix, we will refer to this set as "favorable candidate coatings".  Most notable is that the limiting factor on the anode coatings is the adhesion between the electrolyte and the coating. We hypothesize that as these coatings are more electrochemically stable against Li, their surfaces are also more stable and therefore bond less strongly with the electrolyte. Materials listed in the appendix associated with the Li-metal electrode were only predicted to be well-adhering to Li-metal, not the respective electrolyte, but have an $E_{rxn} = 0~eV$ with Li-metal and meet our other requirements. For this work we focus on the materials which meet all requirements outlined in our methods. We additionally analyze these materials through the lens of ionic conductivity, another necessary metric for high-performance of these coating materials. With the electrolyte materials all exhibiting ionic conductivities $> 10^{-4}~S/cm$, the two barriers to a fast conducting system are the interfacial transfer between electrolyte-coating and electrode-coating, and the ionic conductivity of the coating. The adhesion parameter is designed to mitigate the former, therefore we aim to find coatings with ionic conductivities $> 10^{-4} S/cm$ as well, to mitigate the latter.

Across the three electrolyte systems, sulfides show candidates with more negative (more favorable) adhesion parameters, explained by the upper bounded approximation in the adhesion parameter. If the surfaces used for the adhesion parameter calculation are much more unstable than the most naturally occurring terminations, the electrolyte will appear to have worse adhesion with all possible coating candidates. There are likely more coatings which would adhere well to the electrolyte systems, but with this work we aim to present only our most viable candidates. Because a coating layer is necessary for the stability of the electrolyte/electrode interface, we believe it is worth investigating chemical optimization of the discussed compounds to increase the conductivity, as they meet all other criteria. Below we highlight key insights from our results including: effect anion group on ESW, oxidation state and covalency effects on adhesion, stoichiometric effects within chemical systems, and current literature on our most promising candidates. 

\subsection{Anion Composition of Materials with Good Adhesion}
\begin{figure*}

  \centering
\begin{subfigure}[b]{0.45\textwidth}
\centering
\includegraphics[ width=\textwidth]{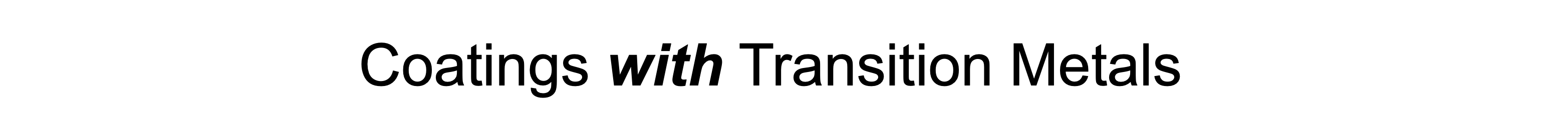}
\label{fig:TMbarslabel}
\end{subfigure}
\hspace{0.3cm}
  \begin{subfigure}[b]{0.45\textwidth}
  \centering
  \includegraphics[ width=\textwidth]{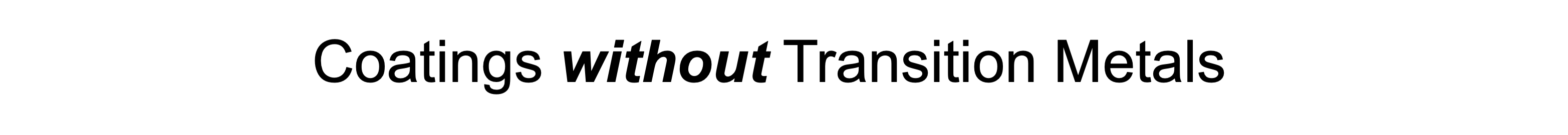}
  \label{fig:barslabel}
  \end{subfigure}
\vfill
    
\begin{subfigure}[b]{0.45\textwidth}
\centering
\includegraphics[ width=\textwidth]{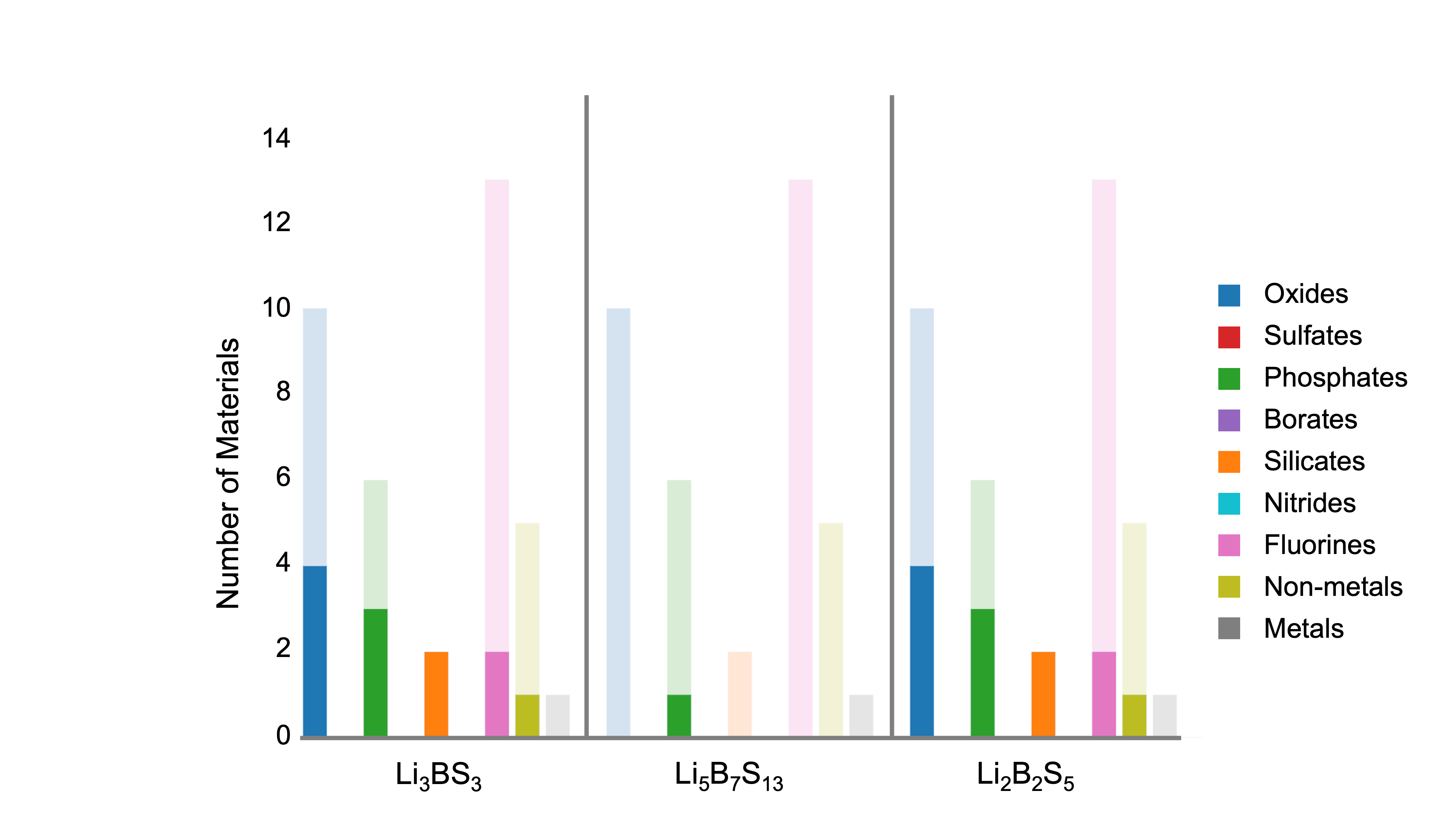}
\label{fig:LBSTMbars}
\end{subfigure}
\hspace{0.3cm}
  \begin{subfigure}[b]{0.45\textwidth}
  \centering
  \includegraphics[ width=\textwidth]{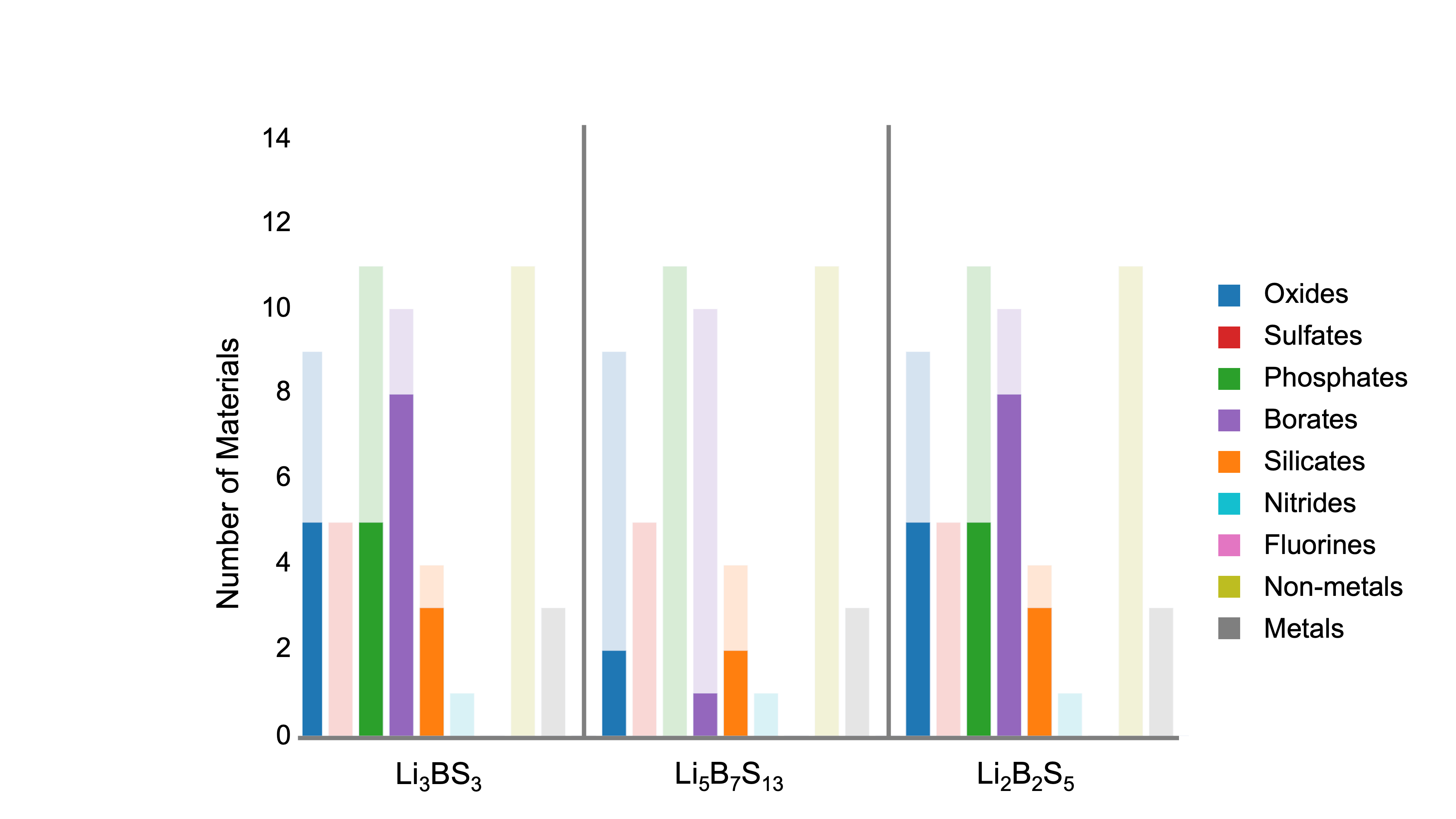}
  \label{fig:LBSbars}
  \end{subfigure}
\vfill

\begin{subfigure}[b]{0.45\textwidth}
\centering
    \includegraphics[width=\textwidth]{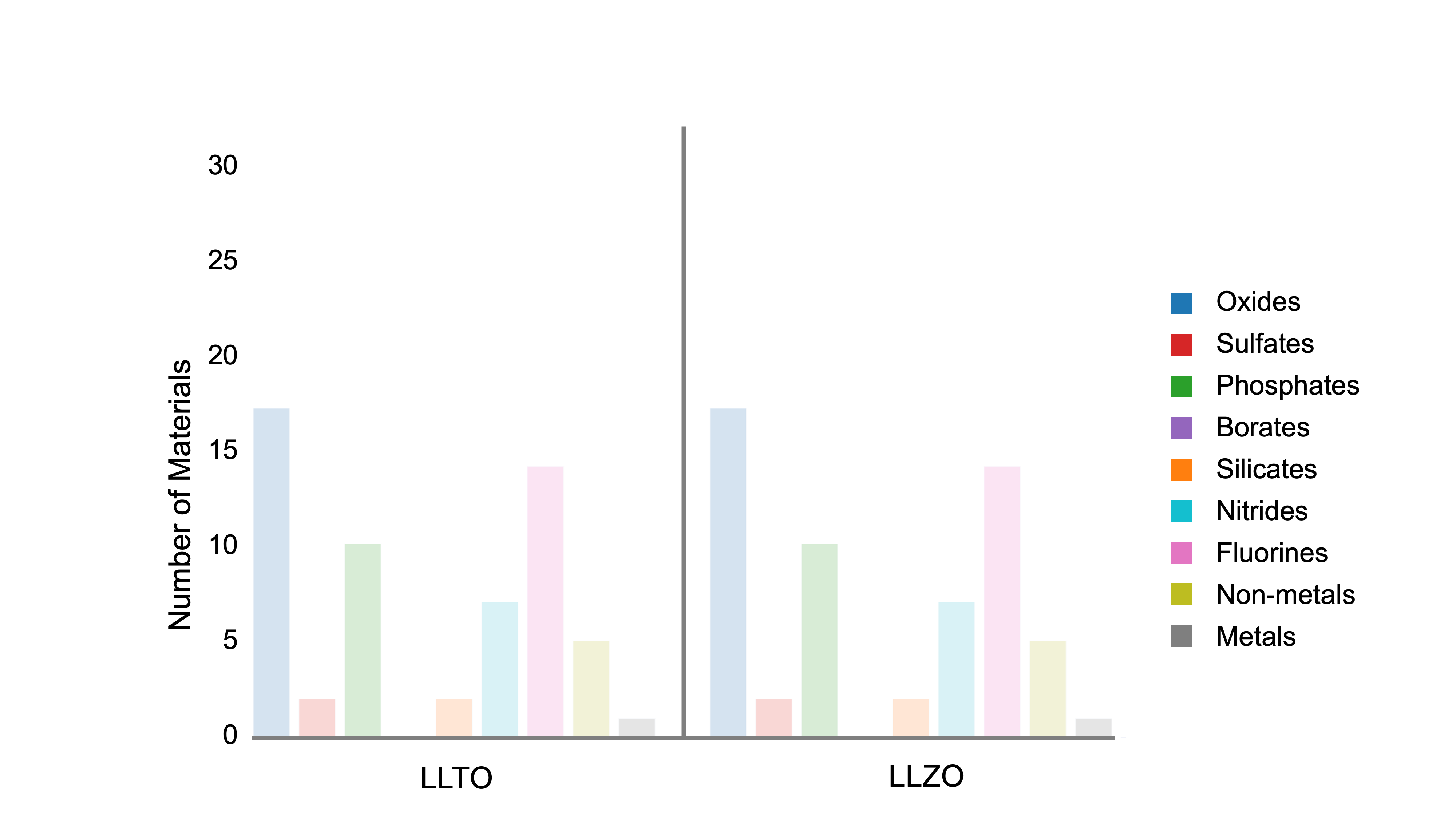}
    \label{fig:LLZOTMbars}
\end{subfigure}
  \hspace{0.3cm}
\begin{subfigure}[b]{0.45\textwidth}
\centering
  \includegraphics[width=\textwidth]{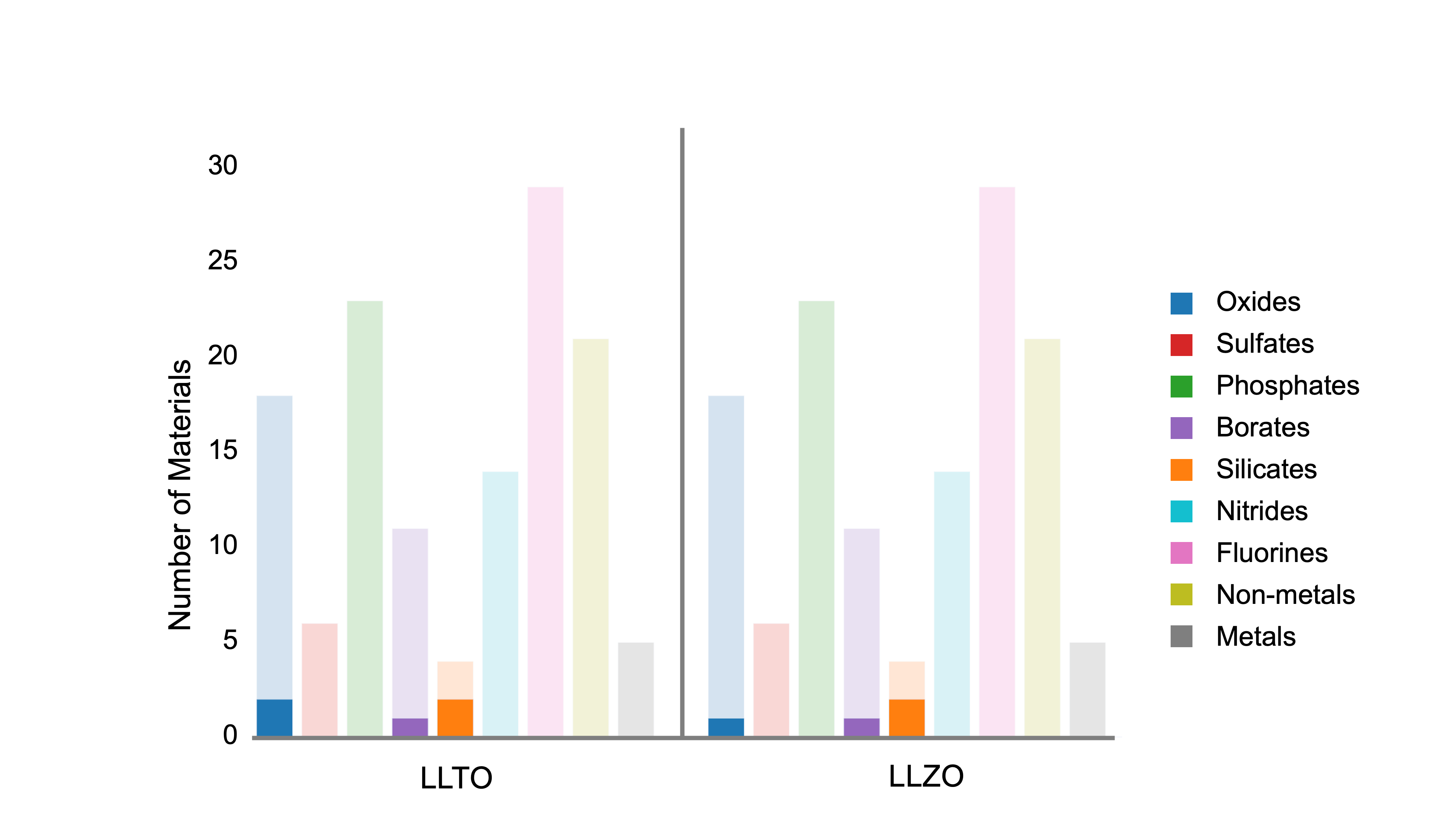}
  \label{fig:LLZObars}
  \end{subfigure}
\vfill

\begin{subfigure}[b]{0.45\textwidth}
\centering
    \includegraphics[ width=\textwidth]{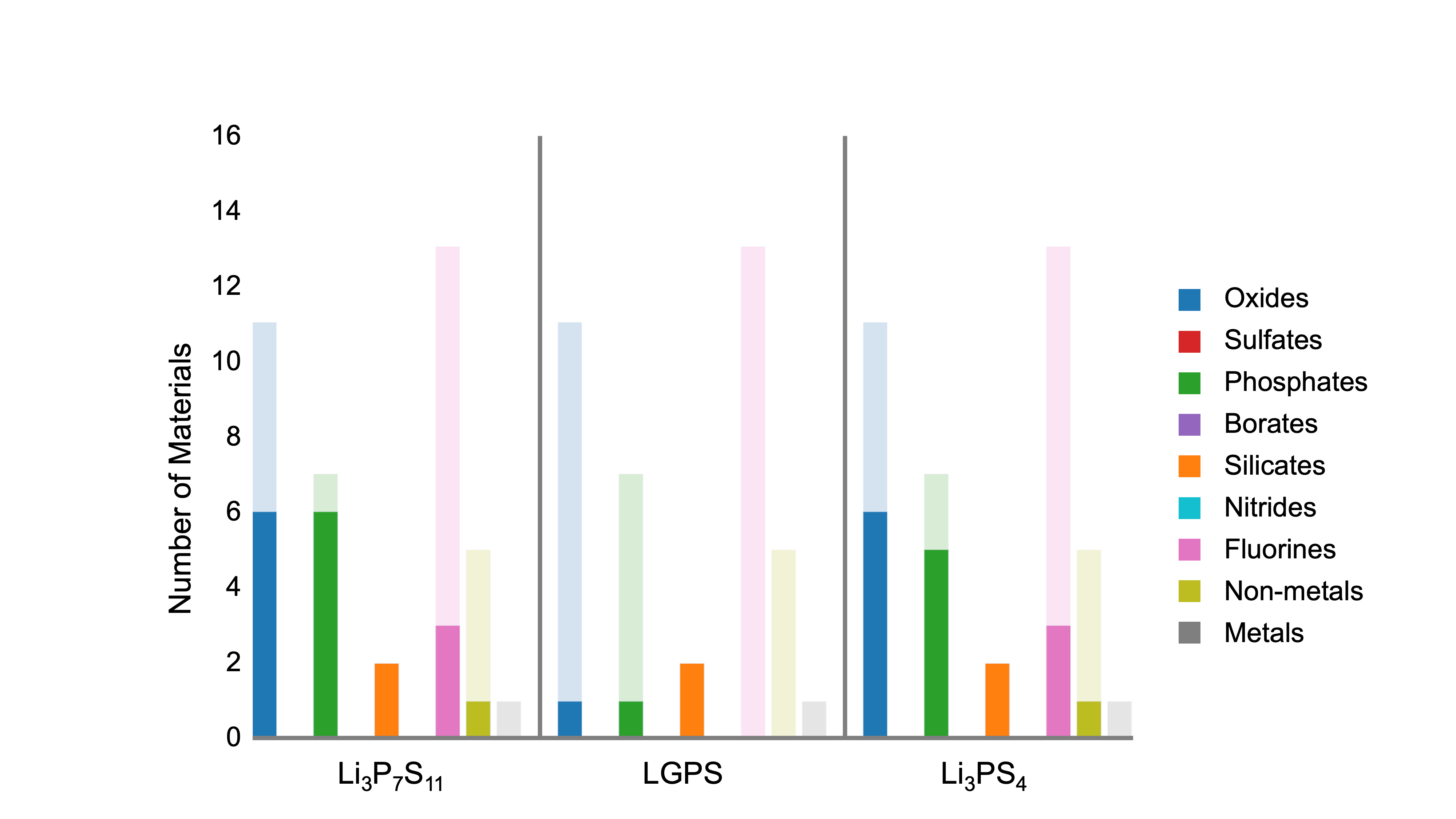}
    \label{fig:LPSTMbars}
\end{subfigure}
  \hspace{0.3cm}
\begin{subfigure}[b]{0.45\textwidth}
\centering
  \includegraphics[ width=\textwidth]{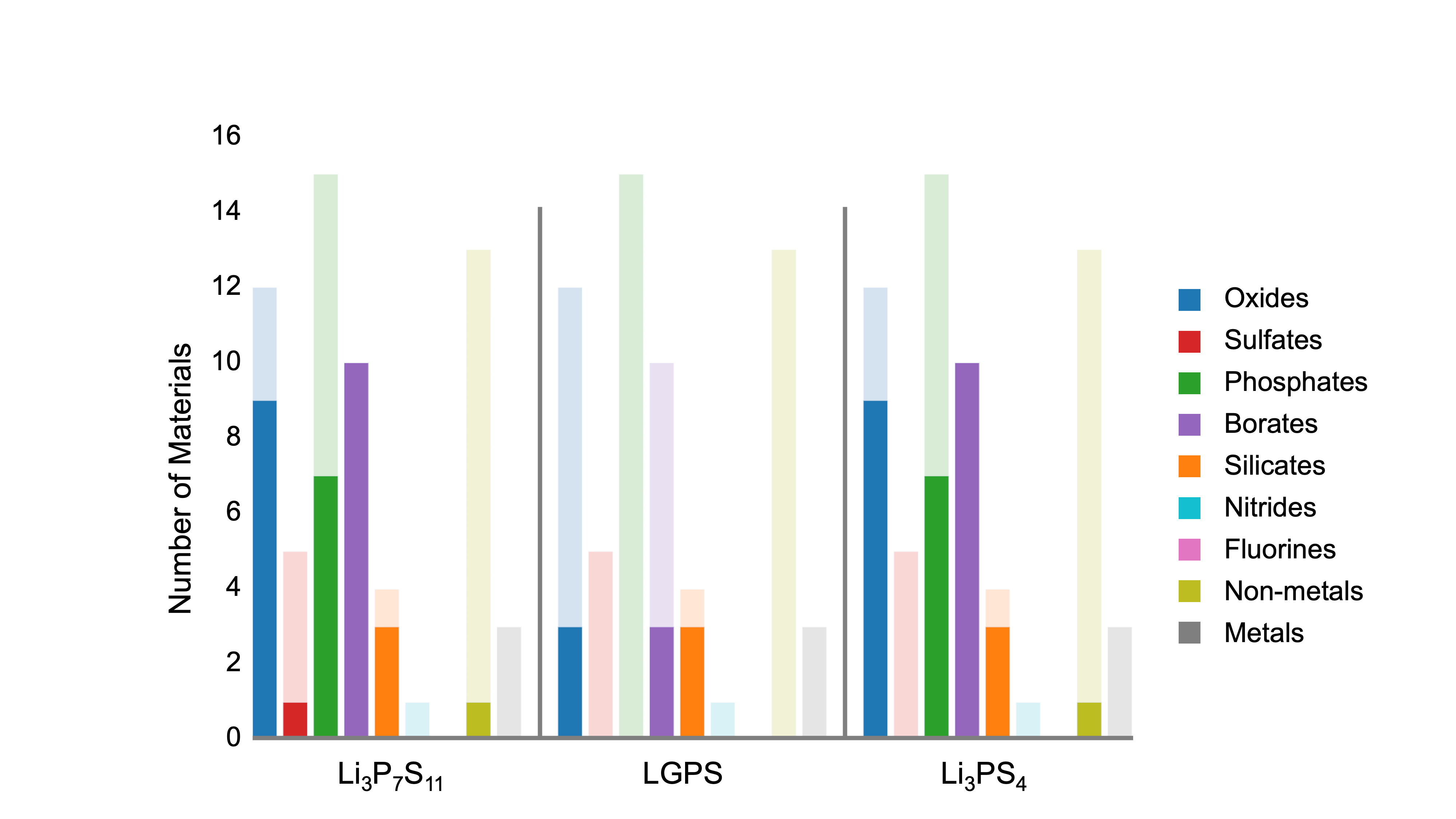}
  \label{fig:LPSbars}
  \end{subfigure}
  \caption{The plots show the coating candidates and their resulting performance sorted by the anionic element or group in the material. The lighter background shows the number of candidates that passed the ESW and elemental screening, and the solid foreground shows the number of materials that have good adhesion with that particular electrolyte.}
\label{fig:bars}
\end{figure*}

We labeled materials by popular anion (or in some cases non-Li) groups to better understand the characteristics of coatings more likely to bond well with electrolytes. The background bars in figure \ref{fig:bars} show the distribution of anion types for materials which had compatible ESWs with each electrolyte group and passed the metrics outlined in section \ref{screening}. 

The most notable difference between electrolyte groups is there are zero non-metal, sulfates, or phosphate coatings available to LLXO systems, which are available to LPS or LBS groups.  The minimum, median, and maximum cleavage energy for each system is listed in table \ref{tab:lyte_cleave}, showing the discrepancy in distributions of terminations. The median values for LLTO and LLZO (and $Li_5B_7S_{13}$) are all significantly higher, representing more unstable surfaces. This approximation is an upper bound on what would be an energy weighted average of all terminations, therefore there would likely be more well-adhering coatings. However, our screening approach allows only extremely stable coatings to achieve all of our screening benchmarks, increasing the likelihood for success in further development. 

\begin{table}[h]
    \centering
    \begin{tabular}{|l|lll|}
    \hline
            & \multicolumn{3}{c|}{Cleavage Energies ($eV/$\AA$^2$)}                     \\
Electrolyte & \multicolumn{1}{l}{minimum} & \multicolumn{1}{l}{median} & maximum \\ \hline
$LLZO$        & \multicolumn{1}{l|}{0.093}   & \multicolumn{1}{l|}{0.188}  & 0.279   \\
$LLTO$        & \multicolumn{1}{l|}{0.036}   & \multicolumn{1}{l|}{0.179}  & 0.289   \\ \hline
$LiB_3S_3$      & \multicolumn{1}{l|}{0.023}   & \multicolumn{1}{l|}{0.069}  & 0.137   \\
$Li_5B_7S_{13}$    & \multicolumn{1}{l|}{0.095}   & \multicolumn{1}{l|}{0.167}  & 0.241   \\
$Li_2B_2S_5$     & \multicolumn{1}{l|}{0.008}   & \multicolumn{1}{l|}{0.074}  & 0.151   \\ \hline
$Li_3PS_4$      & \multicolumn{1}{l|}{0.017}   & \multicolumn{1}{l|}{0.056}  & 0.072   \\
$Li_3P_7S_{11}$    & \multicolumn{1}{l|}{0.011}   & \multicolumn{1}{l|}{0.053}  & 0.086   \\
$Li_{10}GeP_2S_{12}$ & \multicolumn{1}{l|}{0.037}   & \multicolumn{1}{l|}{0.064}  & 0.122   \\ \hline
    \end{tabular}
    
    \caption{Listed are the distribution descriptors of cleavage energies for each electrolyte system, showing the varying distributions across electrolytes.}
    \label{tab:lyte_cleave}
\end{table}


From the plots in figure \ref{fig:bars} we can see the difference that stoichiometry alone has on adhesion. $Li_5B_7S_{13}$ has significantly less favorable adhesion candidates than $Li_3BS_3$ and $Li_2B_2S_5$, which means the available terminations have higher surface energies in $Li_5B_7S_{13}$. $Li_5B_7S_{13}$ has lower Li:B and Li:S ratios, suggesting that Boron and Sulfur are less likely to bond when in contact with other surfaces. We see a related trend in the LPS family, where $Li_{10}GeP_2S_{12}$ produces less candidates for adhesion, even though it has a higher Lithium content. Here, the Germanium atom must destabilize the surface, similarly to the transition metals causing unstable surfaces in the LLXO system. Oxides and Silicates perform well across all compounds, which could ease the process for experimental investigation as these families are typically environmentally stable (i.e., air and water). 

A few anion groups were not favorable for most systems. Nitrides cannot meet the larger ESW requirements for sulfur electrolytes, hence there were very few candidates for which we could approximate adhesion.\cite{richards_miara_wang_kim_ceder_2015} There were no metal compounds with favorable adhesion, which is surprising considering the work done on LiH as a successful coating. It was even shown that LiH was able to decrease the presence of Li dentrites in an LiMg anode, due to the inherent electric field between LiH and LiMg.\cite{zhang_ju_xia_yu_2022} However, our adhesion parameter favors high bonding energies between two materials, and it seems that lithium containing metals are too similar to Li-metal to meet our criteria. Additionally, borates only have favorable adhesion in compounds which don't also have a transition metal, and fluorines only have favorable adhesion in compounds including a transition metal. A highly electronegative element such as Fluorine would be more reactive and stabilize a transition metal ion, as compared to a borate group.

The LLXO systems only adhering with silicate, borate, or oxide coatings speaks to the stability of those coating systems. Silicon and Boron are metalloid polyanionic systems, as opposed to the phosphates and sulfates. The electronegativity difference between Oxygen and Boron or Silicon creates more tightly bound anion groups, which then create a host lattice in which the Li resides. Because of the covalent character within the anionic group, terminations with exposed atoms remain stable at many coordinations. We hypothesize that due to charge sharing, there is a lower probability of exposed unshared electrons, increasing the stability of a termination. The non-oxygen anions, Phosphorous and Silicon, have more flexible oxidation states than Oxygen meaning they can perhaps be stabilized more readily in the presence of a charge compensating transition metal, which is also exhibited in our results.

\subsection{Cathode Coatings}

The different candidates between electrolytes shows that the adhesion between the systems is different and the LPS system is not simply a catch-all for coatings. $LiAl_5O_8$ and $LiAlSiO_4$ are the top two candidates across all systems. $LiAl_5O_8$ has been extensively studied in battery operations across various morphologies. The flexibility of synthesis into nanowire composites, sintered thin films and sol-gel coatings with distinct electrochemical improvements gives a wide space for optimization of this compound.\cite{YANG20191034,AL5O8,WU2021100395} As a Nickel-Manganese-Cobalt cathode coating, $LiAl_5O_8$ was found to increase the coulombic activity and capacity retention with a thin $3 nm$ film.\cite{YANG20191034} Though its independent Li-ion conductivity was found to be only $\sim 10^{-6}~S/cm$, its ability to be cast into films under $10 nm$ can lessen the impact of the lower ionic conductivity.\cite{ito_fujiki_yamada} $LiAl_5O_8$ also limited side reactions and chemical degradation of the cathode material. Wang et al. additionally found $LiAl_5O_8$ to suppress Li-metal dendrites in polymer solid-state batteries.\cite{WU2021100395} Its potential for optimization of other electrochemical metrics make $LiAl_5O_8$ extremely promising for further adaptation. Moreover, as it has been verified for synthesis and assembly into batteries experimentally. \cite{Al5O8cond}
$LiAlSiO_4$ is similarly viable for further investigation. It was shown to improve capacity retention. The synthesis methods take advantage of the glassy nature of this material, whose amorphous phase allows for increased ionic conductivity.\cite{JOHNSON1975403} $LiAlSiO_4$ can be prepared and coated using more simple solution and drying techniques, which makes this appealing to adopt further. There is room for optimization of Li content and cathode particle size, as there have been investigations to optimize weight ratios of the coatings as well as increased ionic conductivity with thin films.\cite{SHINICHI2004325,SUN20151464}

LLXO specific candidates are only slight deviations from the universal candidates $LiAl_5O_8$ and $LiAlSiO_4$, The ordered phases of $LiGaSiO_4$ and $LiAlGeO_4$ have trigonal symmetry, though are not layered materials like the trigonal cathodes. This is likely the cause for the decrease in ionic conductivity in our coating candidates, which only reach favorable ionic conductivities approaching $1000 K$.\cite{QUINTANA1987179,Mukai_2016} However, an extreme improvement up to $\sim10^{-5} S/cm$ was found in solid solutions of $Li_4SiO_4$ and $Li_5GaSi_2O_8$, which gives hope to a possible similar improvement with the mixture of the candidate coatings presented in this work. \cite{QUINTANA1989149} With these structures similar to presented candidate $LiAlSiO_4$, it would be worth investigating glassy phases of these coatings for increased ionic conductivity.

In both sulfide electrolyte systems, $Li_2B_6O_9F_2$ was found as a promising coating. This material has been identified as a coating candidate in previous computational studies, which serves as validation of the methods in this work.\cite{BOF} However, the experimental ionic conductivity was measured at $\sim10^{-10}~S/cm$, making it a likely barrier for conduction between cathodes and electrolytes. \cite{pilz} Only the LPS system shows $K_2LiTa_6(PO_8)_3$ as a promising candidate. The effects of substituting larger Potassium or Cesium atoms for Lithium may have overall negative effects on the ionic conductivity, given the larger size of the atoms. The substitution of 50\% or more Lithium atoms would necessitate a new diffusion mechanism for Lithium diffusion, given the likelihood of inaccessibility to vacancies.

\subsection{Li-metal Coatings}

Most presented coatings for sulfide systems are found to adhere well to Li-metal with an $E_{rxn} = 0~eV$, but they have a less favorable adhesion to the electrolytes. Combined incorporation of LiCl and LiBr systems have recently been investigated to increase their ionic conductivity, though previously LiCl alone has been shown to have insufficient ionic conductivity $(\sim10^{-6}~S/cm)$ .\cite{court-castagnet_1993,tsujimura_ito_yoshida_higashiyama_aihara_machida_park_im_2022} Further work would be necessary to investigate if the stability of these combinations also increases. Previous work by Lutz et al. has specifically investigated Lithium chlorides for the purpose of coatings, though candidate $CsLiCl_2$ has not been specifically investigated.\cite{chun_shim_yu_2021} Across a much wider array of cathodes than was screened in this work, the chlorides showed to have an $E_{rxn} < \sim100~meV$, many we believe can be kinetically stabilized. In addition, all of the explored ternary chlorides have a room temperature conductivity of $>10^{-4}~S/cm$, a promising trend for the family of our new compounds. The only candidate presented for LLXO systems, $Li_5SiN_3$, has previously been investigated as both a cathode and solid-electrolyte material.\cite{AlSiO41,AlSiO42} This is exciting as the material is synthesizable and able to be assembled into a test cell, but further investigation would be needed to ensure we do not expose the electrolyte to the electrode potential. 

If we lowered the $2~eV$ band gap requirement, we find $Li_4CrFe_3O_8$ (band gap = $1.85~eV$) is the only coating in our entire screening process to adhere well to an electrolyte for a Li-metal coating, and it is within the LBS system. Lithium Chromium ferrite is a similar compound which has been studied for it's magnetic properties\cite{PATIL201266,Crferr}. This is a layered oxide, similar to the structure of common cathodes, but with Iron and Chromium in their 3+ oxidation state. We believe there won't be too large of a driving force for redox activity, similar to Zirconium in LLZO. Other Iron-Oxide stoichiometries have been studied for electrode materials. It was found that doping $\alpha-Fe_2O_3$ with Chromium improved the rate performance and Lithium ionic conductivity while Chromium doped $\gamma-Fe_2O_3$ improved the cycling performance. \cite{PAN2018270,CrFeO}

\section{Conclusion}

This work was directed to complement the screening literature for solid-state battery materials, by investigating thermodynamic interactions between surfaces of materials. Experimentally measuring surface energies by contact angle requires extreme surface control, and time consuming computational ionic relaxations of numerous surface terminations makes true interfacial energy infeasible to measure for even a small number of systems.\cite{AuSaphh} We proposed and examined a quick to compute metric for adhesion between two crystalline materials which allows for screening on large scales. In utilizing this parameter to screen for materials, we were able to find materials which account for ionic conductivity, stability, and adhesion at interfaces in solid-state batteries. With our surface-dependent metric, we can extend our evaluation of candidates to more accurately identify and recommend materials whose bulk-metrics have been corroborated by previous experiments. The short coating candidate list for electrolyte systems after screening through more than 19,000 materials demonstrates the difficulty in finding materials that meet all requirements. The sulfide electrolyte coating candidates showed promise when looking at attainable glassy structures, which can increase the ionic conductivity over crystalline forms. The LLXO system had many candidates of various oxides, which upon further investigation could be combined to maximize ionic conductivity. Many of the presented coatings have been experimentally synthesized and assembled into batteries as electrolyte coatings, putting them at the optimization stage for further investigation, accelerating their commercial viability. Specifically, $LiAl_5O_8$ and $LiAlSiO_4$ are top candidates across all electrolyte systems. Their flexibile morphologies allow for more simple synthesis methods and diverse avenues for optimization such as Li content. We believe through better adhesion at the atomic level, solid-state batteries can achieve lower interfacial resistance and avoid mechanical delamination. The materials highlighted in this work can serve as a platform for coating optimization.

\section{Appendix}
\subsection{Cleavage Energy Calculations}
The adhesion parameter was calculated between each candidate and its respective electrolyte pairing for every Lithium containing material in the Materials project data base that met all constraints described in section \ref{screening}. Cleavage energies were implemented as surface energies ($\gamma$) in our adhesion parameter calculation. They require calculations on both the bulk material and isolated slabs created from different possible Miller orientations and terminations where stoichiometry can be sustained. The cleavage energy can be calculated from the following equation: 
\begin{equation}\label{ecleave}
    E_{cleavage} = \frac{E_{slab} - n_{bulk}*E_{bulk}}{2*A_{slab}}
\end{equation}

$E_{cleavage}$ is calculated for all slabs of a particular material. $E_{x}$ represents energies from self-consistent field calculations for both slab and bulk geometries. $n_{bulk}$ is the number of bulk unit cells in the slab and $A_{slab}$ represents the surface area of the slab normal to the surface. The cleavage energy is defined, where A and B represent two surfaces of a slab, by $$E_{cleavage} = \frac{\gamma_A +\gamma_B}{2}$$ therefore ... $$2*E_{cleavage} = \gamma_A +\gamma_B.$$ Because $\gamma_A\geq0$ and $\gamma_B\geq0$, $$2*E_{cleavage} \geq \gamma_A,\gamma_B.$$ and we use $2*E_{cleavage}$ as the upper bound for any $\gamma$ for non-symmetric slabs (i.e., where top and bottom surfaces are not equivalent). 

If the slab expresses symmetry of a mirror or glide plane parallel to the surface, or a 2D rotation normal to the surface, the cleavage energy exactly equals the surface energy. Calculating the exact surface energies for all non-symmetric terminations requires surface pourbaix phase diagrams and chemical potentials for all relevant elements, which is cumbersome. Therefore, we choose to use cleavage energies as an upper bound which suffices in our use case. As the adhesion parameter inequality would be most easily satisfied by a low $max(\gamma_{e})$ and high $min(\gamma_{c})$, we want to use these to bound our approximation for electrolyte slab energies. To adhere to this constraint, $min(\gamma)$ values from cleavage energies for the electrolyte materials include considering the lower surface energies from symmetric slabs. However for coating materials we calculate symmetric slabs with the same 2x upper bound factor on cleavage energies, in order to respect the lower bound approximation on adhesion we are calculating.

For each bulk material, surfaces up to a Miller index of 1 were generated using pymatgen's surface module.\cite{jain_ong_hautier_chen,TRAN2016,TRAN201948} Each surface generally has more than one unique surface termination. We developed an algorithm to ensure that we generate all possible \textit{unique} terminations based on the local environment of surface atoms. Details of this algorithm are described in our previous work.\cite{peter} All slabs generated have a minimum thickness of 10 \AA{} and 15 \AA{} of vacuum between periodic slab repetitions in the $c$-direction to preclude interactions between periodic images.  Self Consistent Field DFT calculations were performed using the Projector Augmented Wave pseudopotential implementation of the Vienna Ab Initio Simulation Package, version 5.4.1. In the DFT calculations, electron exchange and correlation effects are described by the Generalized Gradient Approximation (GGA) functional of PBE. Wave functions are expanded in a plane-wave basis set with a kinetic energy cutoff of 520 eV using gaussian smearing of 200 meV,  and electronic relaxation convergence threshold was 0.1 meV. The energy results of these calculations were input as $E_{slab}$ and $E_{bulk}$ in equation \ref{ecleave}.

\subsection{Full list of Satisfactory Candidates}

Here we present the full list of coating candidates which had a negative adhesion parameter and an $E_{rxn} > -0.1~eV$ when interfaced with the electrolytes. All candidates also had a negative adhesion parameter and an $E_{rxn} > -0.1~eV$ for their interface with cathodes. Candidates marked with $*$ had a positive (unfavorable) adhesion parameter $\leq0.05$ with electrolytes, but did have an $E_{rxn} = 0~eV$ and negative adhesion parameter with Li metal. There was only a single candidate with a favorable adhesion parameter and $E_{rxn} = 0~eV$ with Li metal: $Li_4CrFe_3O_8$ in interface with the LBS system, though its band gap was slightly below our threshold. We include the group of materials with the lowest $E_{rxn}$ value for $Li_5La_3Ta_2O_{12}$ because no coatings had an $E_{rxn} > -0.1~eV$.

\begin{table*}
\centering
\begin{tabular}{|r|cc|cc|}
\hline
\multicolumn{5}{|c|}{\textbf{LLXO System}}\\
\hline
&\multicolumn{2}{c|}{$Li_5La_3Ta_2O_{12}$}&\multicolumn{2}{c|}{$Li_7La_3Zr_2O_{12}$}\\
\multicolumn{1}{|c|}{coating} & \makecell{adhesion\\parameter\\(eV/\AA$^2$)}   & \makecell{$E_{rxn}$ \\(eV)}  & \makecell{adhesion\\parameter\\(eV/\AA$^2$)}   & \makecell{$E_{rxn}$ \\(eV)}\\ \hline
$LiGaSiO_4$ & -0.08  & -0.29 & -0.07  & -0.07 \\
$LiAlGeO_4$ & -0.06 & -0.29 & -0.05 & -0.07 \\
$LiAlSiO_4$ & -0.03 & -0.29 & -0.02 & -0.06 \\
$LiGeBO_4$  & -0.03 & -0.29 & -0.02 & -0.08 \\
$LiAl_5O_8$  & -0.00 & -0.29 &            &   \\  
$*Li_5SiN_3$   & 0.01         & -0.29         & 0.02            & -0.05            \\

\hline
\end{tabular}
\end{table*}

\begin{table*}
\centering
\begin{tabular}{|r|cc|cc|cc|}
\hline
\multicolumn{7}{|c|}{\textbf{LBS System}}\\\hline
&\multicolumn{2}{c|}{$Li_3BS_3$}&\multicolumn{2}{c|}{$Li_2B_2S_5$}&\multicolumn{2}{c|}{$Li_5B_7S_{13}$}\\
\multicolumn{1}{|c|}{coating} & \makecell{adhesion\\parameter\\(eV/\AA$^2$)}    & \makecell{$E_{rxn}$ \\(eV)} & \makecell{adhesion\\parameter\\(eV/\AA$^2$)}    & \makecell{$E_{rxn}$ \\(eV)}& \makecell{adhesion\\parameter\\(eV/\AA$^2$)}    & \makecell{$E_{rxn}$ \\(eV)}\\ \hline
$Li_4CrFe_3O_8$       & \textbf{-0.03} & \textbf{0 }      & \textbf{-0.03} & \textbf{0} &   &  \\

$LiAlSiO_4$          & -0.14 & -0.02 & -0.14 & -0.03 & -0.04 & -0.03   \\
$LiAl_5O_8$           & -0.11 & -0.04 & -0.11 & -0.06 & -0.02 & -0.05   \\
$Li_2B_6O_9F_2$ & -0.09 &-0.02& -0.09&-0.01& & \\
$*Li_2S$    & 0.03          & 0.00           & 0.03           & 0.00         & 0.13         & 0.00          \\
$*Li_2Se$   & 0.04          & 0.00           & 0.03           & -0.02        & 0.13         & 0.00          \\
$*Li_2Te$   & 0.04          & -0.01          & 0.04           & -0.08        & 0.14         & -0.01      \\
$*LiCl$    & 0.05          & 0.00           & 0.05           & 0.00         & 0.15         & 0.00          \\
$*LiF$     & 0.05          & 0.00           & 0.05           & 0.00         & 0.15         & 0.00         \\
\hline
\end{tabular}
\end{table*}

\begin{table*}
\centering
\begin{tabular}{|r|cc|cc|cc|}
\hline
\multicolumn{7}{|c|}{\textbf{LPS System}}\\\hline
&\multicolumn{2}{c|}{$Li_7P_3S_{11}$}&\multicolumn{2}{c|}{$Li_3PS_4$}&\multicolumn{2}{c|}{$Li_{10}GeP_2S_{12}$}\\
\multicolumn{1}{|c|}{coating} & \makecell{adhesion\\parameter\\(eV/\AA$^2$)}    & \makecell{$E_{rxn}$ \\(eV)} & \makecell{adhesion\\parameter\\(eV/\AA$^2$)}    & \makecell{$E_{rxn}$ \\(eV)}& \makecell{adhesion\\parameter\\(eV/\AA$^2$)}    & \makecell{$E_{rxn}$ \\(eV)}\\ \hline
$LiAlSiO_4$          & -0.15 & -0.03 & -0.15 & 0.00  & -0.09 & -0.02 \\
$LiAl_5O_8$           & -0.13 & -0.03 & -0.12 & 0.00  & -0.07 & -0.02 \\
$K_2LiTa_6(PO_8)_3 $    & -0.12 & -0.06 & -0.12 & -0.05 & -0.06 & -0.06 \\
$Li_2B_6O_9F_2$ & -0.11 &-0.03& -0.10&0.00&-0.05 &-0.02 \\
$*Li_2S$    & 0.01          & -0.05          & 0.02           & 0.00         & 0.07         & -0.02         \\
$*Li_2Se$   & 0.02          & -0.04          & 0.02           & 0.00         & 0.08         & -0.02         \\
$*Li_2CN_2$  & 0.02          & -0.08          & 0.03           & -0.06        & 0.09         & -0.06         \\
$*Li_2Te$   & 0.03          & -0.08          & 0.03           & -0.05        & 0.09         & -0.05         \\
$*LiF$     & 0.03          & -0.03          & 0.04           & 0.00         & 0.09         & -0.02         \\
$*LiCl$    & 0.03          & -0.03          & 0.04           & 0.00         & 0.10         & -0.02         \\
$*LiBr$    & 0.04          & -0.03          & 0.04           & 0.00         & 0.10         & -0.02         \\
$*LiI$     & 0.04          & -0.03          & 0.05           & 0.00         & 0.10         & -0.02         \\
$*CsLiCl_2$ & 0.04          & -0.04          & 0.05           & 0.00         & 0.10         & -0.02         \\
$*KLiTe$   & 0.04          & -0.09          & 0.05           & -0.05        & 0.10         & -0.06      \\
\hline
\end{tabular}
\end{table*}

\subsection{Coating placement within the Battery}
\begin{figure}
    \centering
    \includegraphics[width = 0.45\textwidth]{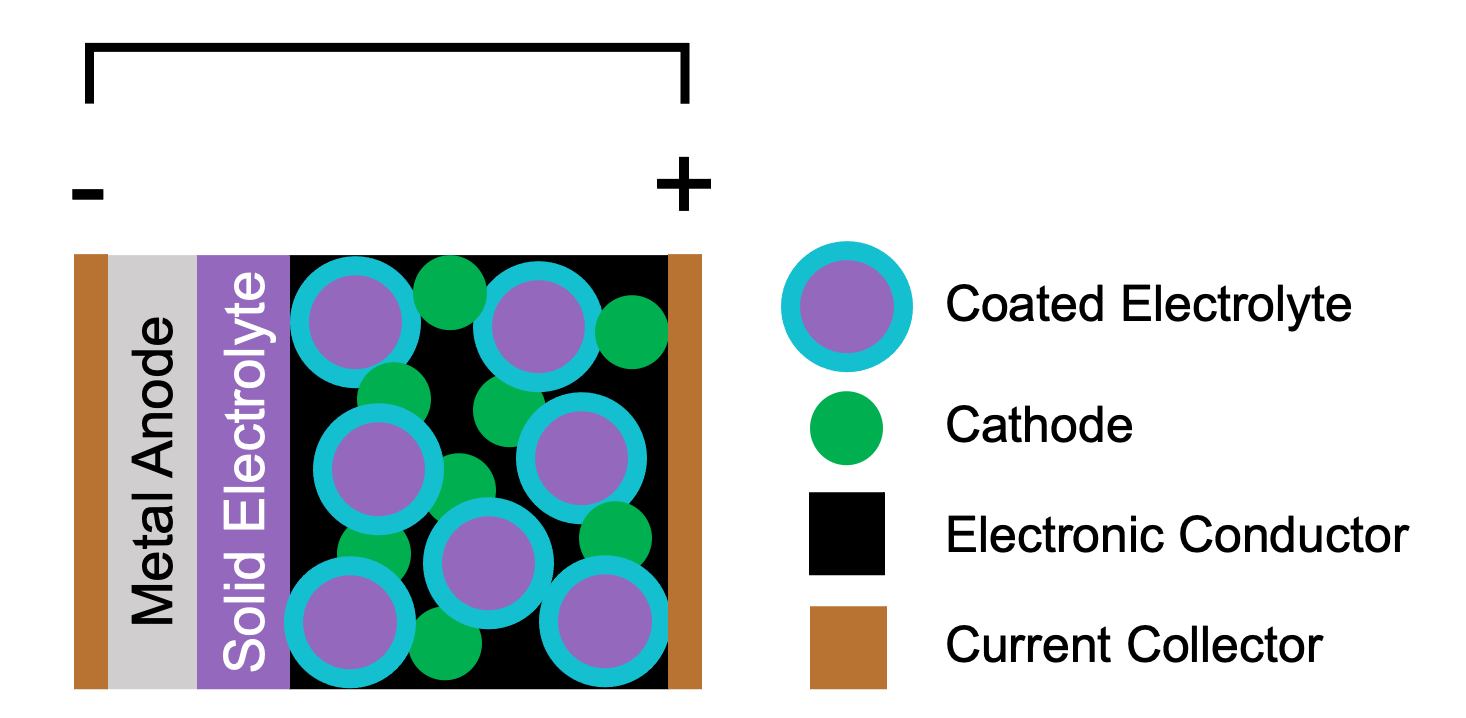}
    \caption{Diagram of solid-state battery.}
    \label{fig:ssb}
\end{figure}

Together with the electrolyte, the coatings help span the electrochemical window between the cathode and the anode, as neither the electrolyte nor the coatings can do it alone. Here we assume the model of a solid-state battery depicted in figure \ref{fig:ssb}, adapted from Hatzell et al's discussion on manufacturing coatings on solid-state batteries.\cite{hatzell_zheng_2021} 

Because our coatings are an extension of the electrolyte, they do not conduct electronically, meaning we need to intentionally place the coating to not interfere with the electronic pathways created by connecting cathode material. Because of this we imagine the electrolyte materials being coated before they undergo processing or combination with the cathode material. With sufficient coating ionic conductivity, the combination of coating and electrolyte extends the operational electrochemical window of the battery. 

The coatings between the anode and solid electrolyte can be applied as a layer.

\section*{Author Contributions}
Brandi Ransom: Methodology, Software, Validation, Formal Analysis, Investigation, Data Curation, Writing - Original Draft, Visualization, Project administration. Akash Ramdas: Writing - Review and Editing, Methodology. Eder Lomeli: Software. Jad Fidawi: Software. Austin Sendek: Writing - Review and Editing, Thomas Devereaux: Writing - Review and Editing, Evan J Reed: Conceptualization, Methodology, Resources. Peter Schindler: Software, Writing - Review and Editing, Supervision. 

\section*{Conflicts of interest}
There are no conflicts to declare.

\section*{Acknowledgements}
Author and Professor Evan J. Reed passed away March 2022. He developed the theory in this work and contributed great guidance towards the project. We are incredibly grateful for his contributions. 



\balance


\bibliography{rsc} 

\providecommand*{\mcitethebibliography}{\thebibliography}
\csname @ifundefined\endcsname{endmcitethebibliography}
{\let\endmcitethebibliography\endthebibliography}{}
\begin{mcitethebibliography}{61}
\providecommand*{\natexlab}[1]{#1}
\providecommand*{\mciteSetBstSublistMode}[1]{}
\providecommand*{\mciteSetBstMaxWidthForm}[2]{}
\providecommand*{\mciteBstWouldAddEndPuncttrue}
  {\def\EndOfBibitem{\unskip.}}
\providecommand*{\mciteBstWouldAddEndPunctfalse}
  {\let\EndOfBibitem\relax}
\providecommand*{\mciteSetBstMidEndSepPunct}[3]{}
\providecommand*{\mciteSetBstSublistLabelBeginEnd}[3]{}
\providecommand*{\EndOfBibitem}{}
\mciteSetBstSublistMode{f}
\mciteSetBstMaxWidthForm{subitem}
{(\emph{\alph{mcitesubitemcount}})}
\mciteSetBstSublistLabelBeginEnd{\mcitemaxwidthsubitemform\space}
{\relax}{\relax}

\bibitem[bat()]{batterymarket}
\emph{Battery Market Size; Share: Industry Report, 2020-2027},
  \url{https://www.grandviewresearch.com/industry-analysis/battery-market}\relax
\mciteBstWouldAddEndPuncttrue
\mciteSetBstMidEndSepPunct{\mcitedefaultmidpunct}
{\mcitedefaultendpunct}{\mcitedefaultseppunct}\relax
\EndOfBibitem
\bibitem[Lv \emph{et~al.}(2022)Lv, Zhou, Zhong, Yan, Srinivasan, Seh, Liu, Pan,
  Li, Wen, and Yan]{matdisc1}
C.~Lv, X.~Zhou, L.~Zhong, C.~Yan, M.~Srinivasan, Z.~W. Seh, C.~Liu, H.~Pan,
  S.~Li, Y.~Wen and Q.~Yan, \emph{Advanced Materials}, 2022, \textbf{34},
  2101474\relax
\mciteBstWouldAddEndPuncttrue
\mciteSetBstMidEndSepPunct{\mcitedefaultmidpunct}
{\mcitedefaultendpunct}{\mcitedefaultseppunct}\relax
\EndOfBibitem
\bibitem[Liu \emph{et~al.}(2020)Liu, Guo, Zou, Li, and Shi]{matdisc2}
Y.~Liu, B.~Guo, X.~Zou, Y.~Li and S.~Shi, \emph{Energy Storage Materials},
  2020, \textbf{31}, 434--450\relax
\mciteBstWouldAddEndPuncttrue
\mciteSetBstMidEndSepPunct{\mcitedefaultmidpunct}
{\mcitedefaultendpunct}{\mcitedefaultseppunct}\relax
\EndOfBibitem
\bibitem[Sendek \emph{et~al.}(2022)Sendek, Cubuk, Ransom, Nanda, and
  Reed]{matdisc3}
A.~D. Sendek, E.~D. Cubuk, B.~Ransom, J.~Nanda and E.~J. Reed, in
  \emph{Machine-Learning and Data-Intensive Methods for Accelerating the
  Development of Rechargeable Battery Chemistries: A Review}, John Wiley Sons,
  Ltd, 2022, ch.~16, pp. 393--409\relax
\mciteBstWouldAddEndPuncttrue
\mciteSetBstMidEndSepPunct{\mcitedefaultmidpunct}
{\mcitedefaultendpunct}{\mcitedefaultseppunct}\relax
\EndOfBibitem
\bibitem[Shapshak(2019)]{shapshak_2019}
T.~Shapshak, \emph{Cobalt Lawsuit Against Tech Giants Over Child Labour A
  'Global Flashpoint Of Corporate Social Responsibility'}, 2019,
  Forbes.com\relax
\mciteBstWouldAddEndPuncttrue
\mciteSetBstMidEndSepPunct{\mcitedefaultmidpunct}
{\mcitedefaultendpunct}{\mcitedefaultseppunct}\relax
\EndOfBibitem
\bibitem[Indu \emph{et~al.}(2021)Indu, Alexander, Sreejith, Abraham, and
  Murugan]{INDU2021100804}
M.~Indu, G.~Alexander, O.~Sreejith, S.~Abraham and R.~Murugan, \emph{Materials
  Today Energy}, 2021, \textbf{21}, 100804\relax
\mciteBstWouldAddEndPuncttrue
\mciteSetBstMidEndSepPunct{\mcitedefaultmidpunct}
{\mcitedefaultendpunct}{\mcitedefaultseppunct}\relax
\EndOfBibitem
\bibitem[Xiao \emph{et~al.}(2019)Xiao, Wang, Bo, Kim, Miara, and
  Ceder]{xiao_wang_bo_kim_miara_ceder_2019}
Y.~Xiao, Y.~Wang, S.-H. Bo, J.~C. Kim, L.~J. Miara and G.~Ceder, \emph{Nature
  Reviews Materials}, 2019, \textbf{5}, 105–126\relax
\mciteBstWouldAddEndPuncttrue
\mciteSetBstMidEndSepPunct{\mcitedefaultmidpunct}
{\mcitedefaultendpunct}{\mcitedefaultseppunct}\relax
\EndOfBibitem
\bibitem[Camacho-Forero and Balbuena(2018)]{CAMACHOFORERO2018782}
L.~E. Camacho-Forero and P.~B. Balbuena, \emph{Journal of Power Sources}, 2018,
  \textbf{396}, 782--790\relax
\mciteBstWouldAddEndPuncttrue
\mciteSetBstMidEndSepPunct{\mcitedefaultmidpunct}
{\mcitedefaultendpunct}{\mcitedefaultseppunct}\relax
\EndOfBibitem
\bibitem[Zhu \emph{et~al.}(2016)Zhu, He, and Mo]{zhu_he_mo_2016}
Y.~Zhu, X.~He and Y.~Mo, \emph{Journal of Materials Chemistry A}, 2016,
  \textbf{4}, 3253–3266\relax
\mciteBstWouldAddEndPuncttrue
\mciteSetBstMidEndSepPunct{\mcitedefaultmidpunct}
{\mcitedefaultendpunct}{\mcitedefaultseppunct}\relax
\EndOfBibitem
\bibitem[Kim and Rupp(2020)]{kim_rupp_2020}
K.~J. Kim and J.~L. Rupp, \emph{Energy amp; Environmental Science}, 2020,
  \textbf{13}, 4930–4945\relax
\mciteBstWouldAddEndPuncttrue
\mciteSetBstMidEndSepPunct{\mcitedefaultmidpunct}
{\mcitedefaultendpunct}{\mcitedefaultseppunct}\relax
\EndOfBibitem
\bibitem[Jiang \emph{et~al.}(2019)Jiang, Han, Wang, and
  Wang]{jiang_han_wang_wang_2019}
Z.~Jiang, Q.~Han, S.~Wang and H.~Wang, \emph{ChemElectroChem}, 2019,
  \textbf{6}, 2970–2983\relax
\mciteBstWouldAddEndPuncttrue
\mciteSetBstMidEndSepPunct{\mcitedefaultmidpunct}
{\mcitedefaultendpunct}{\mcitedefaultseppunct}\relax
\EndOfBibitem
\bibitem[Schlenker \emph{et~al.}(2020)Schlenker, Stępień, Koch, Hupfer,
  Indris, Roling, Miß, Fuchs, Wilhelmi, and Ehrenberg]{schlenker}
R.~Schlenker, D.~Stępień, P.~Koch, T.~Hupfer, S.~Indris, B.~Roling, V.~Miß,
  A.~Fuchs, M.~Wilhelmi and H.~Ehrenberg, \emph{ACS Applied Materials \
  Interfaces}, 2020, \textbf{12}, 20012--20025\relax
\mciteBstWouldAddEndPuncttrue
\mciteSetBstMidEndSepPunct{\mcitedefaultmidpunct}
{\mcitedefaultendpunct}{\mcitedefaultseppunct}\relax
\EndOfBibitem
\bibitem[Zhou \emph{et~al.}(2022)Zhou, Zhang, Shen, Fang, Kong, Feng, Xie,
  Wang, Hu, Wang, and
  et~al.]{zhou_zhang_shen_fang_kong_feng_xie_wang_hu_wang_etal._2022}
X.~Zhou, Y.~Zhang, M.~Shen, Z.~Fang, T.~Kong, W.~Feng, Y.~Xie, F.~Wang, B.~Hu,
  Y.~Wang and et~al., \emph{Advanced Energy Materials}, 2022, \textbf{12},
  2103932\relax
\mciteBstWouldAddEndPuncttrue
\mciteSetBstMidEndSepPunct{\mcitedefaultmidpunct}
{\mcitedefaultendpunct}{\mcitedefaultseppunct}\relax
\EndOfBibitem
\bibitem[Fan \emph{et~al.}(2018)Fan, Ji, Han, Yue, Chen, Chen, Deng, Jiang, and
  Wang]{fan_ji_han_yue_chen_chen_deng_jiang_wang_2018}
X.~Fan, X.~Ji, F.~Han, J.~Yue, J.~Chen, L.~Chen, T.~Deng, J.~Jiang and C.~Wang,
  \emph{Science Advances}, 2018, \textbf{4}, \relax
\mciteBstWouldAddEndPuncttrue
\mciteSetBstMidEndSepPunct{\mcitedefaultmidpunct}
{\mcitedefaultendpunct}{\mcitedefaultseppunct}\relax
\EndOfBibitem
\bibitem[Connell \emph{et~al.}(2020)Connell, Fuchs, Hartmann, Krauskopf, Zhu,
  Sann, Garcia-Mendez, Sakamoto, Tepavcevic, Janek, and
  et~al.]{connell_fuchs_hartmann_krauskopf_zhu_sann_garcia-mendez_sakamoto_tepavcevic_janek_etal._2020}
J.~G. Connell, T.~Fuchs, H.~Hartmann, T.~Krauskopf, Y.~Zhu, J.~Sann,
  R.~Garcia-Mendez, J.~Sakamoto, S.~Tepavcevic, J.~Janek and et~al.,
  \emph{Chemistry of Materials}, 2020, \textbf{32}, 10207–10215\relax
\mciteBstWouldAddEndPuncttrue
\mciteSetBstMidEndSepPunct{\mcitedefaultmidpunct}
{\mcitedefaultendpunct}{\mcitedefaultseppunct}\relax
\EndOfBibitem
\bibitem[Zisman(1964)]{contact_angle}
F.~Zisman, W., \emph{ACS}, 1964,  1--51\relax
\mciteBstWouldAddEndPuncttrue
\mciteSetBstMidEndSepPunct{\mcitedefaultmidpunct}
{\mcitedefaultendpunct}{\mcitedefaultseppunct}\relax
\EndOfBibitem
\bibitem[Ehsani \emph{et~al.}(2021)Ehsani, Boyd, Wang, and Grady]{pete1}
H.~Ehsani, J.~D. Boyd, J.~Wang and M.~E. Grady, \emph{Applied Mechanics
  Reviews}, 2021, \textbf{73}, \relax
\mciteBstWouldAddEndPuncttrue
\mciteSetBstMidEndSepPunct{\mcitedefaultmidpunct}
{\mcitedefaultendpunct}{\mcitedefaultseppunct}\relax
\EndOfBibitem
\bibitem[Wolloch \emph{et~al.}(2022)Wolloch, Losi, Chehaimi, Yalcin, Ferrario,
  and Righi]{WOLLOCH2022111302}
M.~Wolloch, G.~Losi, O.~Chehaimi, F.~Yalcin, M.~Ferrario and M.~C. Righi,
  \emph{Computational Materials Science}, 2022, \textbf{207}, 111302\relax
\mciteBstWouldAddEndPuncttrue
\mciteSetBstMidEndSepPunct{\mcitedefaultmidpunct}
{\mcitedefaultendpunct}{\mcitedefaultseppunct}\relax
\EndOfBibitem
\bibitem[Wolloch \emph{et~al.}(2019)Wolloch, Losi, and Ferrario]{pete2}
M.~Wolloch, G.~Losi and M.~Ferrario, \emph{Sci Rep}, 2019, \textbf{9}, \relax
\mciteBstWouldAddEndPuncttrue
\mciteSetBstMidEndSepPunct{\mcitedefaultmidpunct}
{\mcitedefaultendpunct}{\mcitedefaultseppunct}\relax
\EndOfBibitem
\bibitem[Jain \emph{et~al.}(2013)Jain, Ong, Hautier, Chen, Richards, Dacek,
  Cholia, Gunter, Skinner, Ceder, and et~al.]{jain_ong_hautier_chen}
A.~Jain, S.~P. Ong, G.~Hautier, W.~Chen, W.~D. Richards, S.~Dacek, S.~Cholia,
  D.~Gunter, D.~Skinner, G.~Ceder and et~al., \emph{APL Materials}, 2013,
  \textbf{1}, 011002\relax
\mciteBstWouldAddEndPuncttrue
\mciteSetBstMidEndSepPunct{\mcitedefaultmidpunct}
{\mcitedefaultendpunct}{\mcitedefaultseppunct}\relax
\EndOfBibitem
\bibitem[Ong \emph{et~al.}(2015)Ong, Cholia, Jain, Brafman, Gunter, Ceder, and
  Persson]{MP_api}
Ong, Cholia, Jain, Brafman, Gunter, Ceder and Persson, \emph{Computational
  Materials Science}, 2015, \textbf{97}, \relax
\mciteBstWouldAddEndPuncttrue
\mciteSetBstMidEndSepPunct{\mcitedefaultmidpunct}
{\mcitedefaultendpunct}{\mcitedefaultseppunct}\relax
\EndOfBibitem
\bibitem[Jain \emph{et~al.}(2011)Jain, Hautier, Ong, Moore, Fischer, Persson,
  and Ceder]{PhysRevB.84.045115}
A.~Jain, G.~Hautier, S.~P. Ong, C.~J. Moore, C.~C. Fischer, K.~A. Persson and
  G.~Ceder, \emph{Phys. Rev. B}, 2011, \textbf{84}, 045115\relax
\mciteBstWouldAddEndPuncttrue
\mciteSetBstMidEndSepPunct{\mcitedefaultmidpunct}
{\mcitedefaultendpunct}{\mcitedefaultseppunct}\relax
\EndOfBibitem
\bibitem[Zaghib \emph{et~al.}(2012)Zaghib, Dubé, Dallaire, Galoustov, Guerfi,
  Ramanathan, Benmayza, Prakash, Mauger, Julien, and
  et~al.]{zaghib_dub_dallaire_galoustov_guerfi_ramanathan_benmayza_prakash_mauger_julien_etal._2012}
K.~Zaghib, J.~Dubé, A.~Dallaire, K.~Galoustov, A.~Guerfi, M.~Ramanathan,
  A.~Benmayza, J.~Prakash, A.~Mauger, C.~Julien and et~al., \emph{Journal of
  Power Sources}, 2012, \textbf{219}, 36–44\relax
\mciteBstWouldAddEndPuncttrue
\mciteSetBstMidEndSepPunct{\mcitedefaultmidpunct}
{\mcitedefaultendpunct}{\mcitedefaultseppunct}\relax
\EndOfBibitem
\bibitem[Kalluri \emph{et~al.}(2017)Kalluri, Yoon, Jo, Park, Myeong, Kim, Dou,
  Guo, and Cho]{kalluri_yoon_jo_park_myeong_kim_dou_guo_cho_2017}
S.~Kalluri, M.~Yoon, M.~Jo, S.~Park, S.~Myeong, J.~Kim, S.~X. Dou, Z.~Guo and
  J.~Cho, \emph{Advanced Energy Materials}, 2017, \textbf{7}, 1601507\relax
\mciteBstWouldAddEndPuncttrue
\mciteSetBstMidEndSepPunct{\mcitedefaultmidpunct}
{\mcitedefaultendpunct}{\mcitedefaultseppunct}\relax
\EndOfBibitem
\bibitem[Ding \emph{et~al.}(2010)Ding, Xie, Cao, Zhu, Yu, and
  Zhao]{ding_xie_cao_zhu_yu_zhao_2010}
Y.-L. Ding, J.~Xie, G.-S. Cao, T.-J. Zhu, H.-M. Yu and X.-B. Zhao,
  \emph{Advanced Functional Materials}, 2010, \textbf{21}, 348–355\relax
\mciteBstWouldAddEndPuncttrue
\mciteSetBstMidEndSepPunct{\mcitedefaultmidpunct}
{\mcitedefaultendpunct}{\mcitedefaultseppunct}\relax
\EndOfBibitem
\bibitem[Park \emph{et~al.}(2021)Park, Zhu, Torres‐Castanedo, Jung, Luu,
  Kahvecioglu, Yoo, Seo, Downing, Lim, and et~al.]{park_zhu_torres}
K.~Park, Y.~Zhu, C.~G. Torres‐Castanedo, H.~J. Jung, N.~S. Luu,
  O.~Kahvecioglu, Y.~Yoo, J.~T. Seo, J.~R. Downing, H.~Lim and et~al.,
  \emph{Advanced Materials}, 2021, \textbf{34}, 2106402\relax
\mciteBstWouldAddEndPuncttrue
\mciteSetBstMidEndSepPunct{\mcitedefaultmidpunct}
{\mcitedefaultendpunct}{\mcitedefaultseppunct}\relax
\EndOfBibitem
\bibitem[Schwietert \emph{et~al.}(2021)Schwietert, Vasileiadis, and
  Wagemaker]{schwietert_vasileiadis_wagemaker_2021}
T.~K. Schwietert, A.~Vasileiadis and M.~Wagemaker, \emph{JACS Au}, 2021,
  \textbf{1}, 1488–1496\relax
\mciteBstWouldAddEndPuncttrue
\mciteSetBstMidEndSepPunct{\mcitedefaultmidpunct}
{\mcitedefaultendpunct}{\mcitedefaultseppunct}\relax
\EndOfBibitem
\bibitem[Ong \emph{et~al.}(2008)Ong, Wang, Kang, and
  Ceder]{ong_wang_kang_ceder_2008}
S.~P. Ong, L.~Wang, B.~Kang and G.~Ceder, \emph{Chemistry of Materials}, 2008,
  \textbf{20}, 1798–1807\relax
\mciteBstWouldAddEndPuncttrue
\mciteSetBstMidEndSepPunct{\mcitedefaultmidpunct}
{\mcitedefaultendpunct}{\mcitedefaultseppunct}\relax
\EndOfBibitem
\bibitem[Han \emph{et~al.}(2016)Han, Zhu, He, Mo, and
  Wang]{han_zhu_he_mo_wang_2016}
F.~Han, Y.~Zhu, X.~He, Y.~Mo and C.~Wang, \emph{Advanced Energy Materials},
  2016, \textbf{6}, 1501590\relax
\mciteBstWouldAddEndPuncttrue
\mciteSetBstMidEndSepPunct{\mcitedefaultmidpunct}
{\mcitedefaultendpunct}{\mcitedefaultseppunct}\relax
\EndOfBibitem
\bibitem[Sendek \emph{et~al.}(2020)Sendek, Antoniuk, Cubuk, Ransom, Francisco,
  Buettner-Garrett, Cui, and
  Reed]{sendek_antoniuk_cubuk_ransom_francisco_buettner-garrett_cui_reed_2020}
A.~D. Sendek, E.~R. Antoniuk, E.~D. Cubuk, B.~Ransom, B.~E. Francisco,
  J.~Buettner-Garrett, Y.~Cui and E.~J. Reed, \emph{ACS Applied Materials amp;
  Interfaces}, 2020, \textbf{12}, 37957–37966\relax
\mciteBstWouldAddEndPuncttrue
\mciteSetBstMidEndSepPunct{\mcitedefaultmidpunct}
{\mcitedefaultendpunct}{\mcitedefaultseppunct}\relax
\EndOfBibitem
\bibitem[Rong \emph{et~al.}(2016)Rong, Kitchaev, Canepa, Huang, and
  Ceder]{pymatgen}
Z.~Rong, D.~Kitchaev, P.~Canepa, W.~Huang and G.~Ceder, \emph{The Journal of
  Chemical Physics}, 2016, \textbf{145}, 074112\relax
\mciteBstWouldAddEndPuncttrue
\mciteSetBstMidEndSepPunct{\mcitedefaultmidpunct}
{\mcitedefaultendpunct}{\mcitedefaultseppunct}\relax
\EndOfBibitem
\bibitem[Winterbottom(1967)]{WINTERBOTTOM1967303}
W.~Winterbottom, \emph{Acta Metallurgica}, 1967, \textbf{15}, 303--310\relax
\mciteBstWouldAddEndPuncttrue
\mciteSetBstMidEndSepPunct{\mcitedefaultmidpunct}
{\mcitedefaultendpunct}{\mcitedefaultseppunct}\relax
\EndOfBibitem
\bibitem[Richards \emph{et~al.}(2015)Richards, Miara, Wang, Kim, and
  Ceder]{richards_miara_wang_kim_ceder_2015}
W.~D. Richards, L.~J. Miara, Y.~Wang, J.~C. Kim and G.~Ceder, \emph{Chemistry
  of Materials}, 2015, \textbf{28}, 266–273\relax
\mciteBstWouldAddEndPuncttrue
\mciteSetBstMidEndSepPunct{\mcitedefaultmidpunct}
{\mcitedefaultendpunct}{\mcitedefaultseppunct}\relax
\EndOfBibitem
\bibitem[Zhang \emph{et~al.}(2022)Zhang, Ju, Xia, and Yu]{zhang_ju_xia_yu_2022}
H.~Zhang, S.~Ju, G.~Xia and X.~Yu, \emph{Science Advances}, 2022, \textbf{8},
  \relax
\mciteBstWouldAddEndPuncttrue
\mciteSetBstMidEndSepPunct{\mcitedefaultmidpunct}
{\mcitedefaultendpunct}{\mcitedefaultseppunct}\relax
\EndOfBibitem
\bibitem[Yang \emph{et~al.}(2019)Yang, Lin, Guo, Shao, Zhang, Zhang, Yan, and
  Volinsky]{YANG20191034}
F.~Yang, S.~Lin, Z.~Guo, Y.~Shao, B.~Zhang, X.~Zhang, S.~Yan and A.~A.
  Volinsky, \emph{Journal of Alloys and Compounds}, 2019, \textbf{805},
  1034--1043\relax
\mciteBstWouldAddEndPuncttrue
\mciteSetBstMidEndSepPunct{\mcitedefaultmidpunct}
{\mcitedefaultendpunct}{\mcitedefaultseppunct}\relax
\EndOfBibitem
\bibitem[Temeche \emph{et~al.}(2020)Temeche, Indris, and Laine]{AL5O8}
E.~Temeche, S.~Indris and R.~M. Laine, \emph{ACS Applied Materials \
  Interfaces}, 2020, \textbf{12}, 46119--46131\relax
\mciteBstWouldAddEndPuncttrue
\mciteSetBstMidEndSepPunct{\mcitedefaultmidpunct}
{\mcitedefaultendpunct}{\mcitedefaultseppunct}\relax
\EndOfBibitem
\bibitem[Wu \emph{et~al.}(2021)Wu, Lei, and Wang]{WU2021100395}
Y.~Wu, D.~Lei and C.~Wang, \emph{Materials Today Physics}, 2021, \textbf{18},
  100395\relax
\mciteBstWouldAddEndPuncttrue
\mciteSetBstMidEndSepPunct{\mcitedefaultmidpunct}
{\mcitedefaultendpunct}{\mcitedefaultseppunct}\relax
\EndOfBibitem
\bibitem[Ito \emph{et~al.}(2014)Ito, Fujiki, Yamada, Aihara, Park, Kim, Baek,
  Lee, Doo, Machida, and et~al.]{ito_fujiki_yamada}
S.~Ito, S.~Fujiki, T.~Yamada, Y.~Aihara, Y.~Park, T.~Y. Kim, S.-W. Baek, J.-M.
  Lee, S.~Doo, N.~Machida and et~al., \emph{Journal of Power Sources}, 2014,
  \textbf{248}, 943–950\relax
\mciteBstWouldAddEndPuncttrue
\mciteSetBstMidEndSepPunct{\mcitedefaultmidpunct}
{\mcitedefaultendpunct}{\mcitedefaultseppunct}\relax
\EndOfBibitem
\bibitem[Miyakawa \emph{et~al.}(2022)Miyakawa, Matsuda, and
  Tanibata]{Al5O8cond}
Miyakawa, Matsuda and Tanibata, \emph{JSci Rep}, 2022, \textbf{12}, 16672\relax
\mciteBstWouldAddEndPuncttrue
\mciteSetBstMidEndSepPunct{\mcitedefaultmidpunct}
{\mcitedefaultendpunct}{\mcitedefaultseppunct}\relax
\EndOfBibitem
\bibitem[Johnson \emph{et~al.}(1975)Johnson, Morosin, Knotek, and
  Biefeld]{JOHNSON1975403}
R.~Johnson, B.~Morosin, M.~Knotek and R.~Biefeld, \emph{Physics Letters A},
  1975, \textbf{54}, 403--404\relax
\mciteBstWouldAddEndPuncttrue
\mciteSetBstMidEndSepPunct{\mcitedefaultmidpunct}
{\mcitedefaultendpunct}{\mcitedefaultseppunct}\relax
\EndOfBibitem
\bibitem[SHI(2004)]{SHINICHI2004325}
\emph{Solid State Ionics}, 2004, \textbf{167}, 325--329\relax
\mciteBstWouldAddEndPuncttrue
\mciteSetBstMidEndSepPunct{\mcitedefaultmidpunct}
{\mcitedefaultendpunct}{\mcitedefaultseppunct}\relax
\EndOfBibitem
\bibitem[Sun \emph{et~al.}(2015)Sun, Li, Qiao, Cao, Wang, and Ye]{SUN20151464}
Y.~Sun, F.~Li, Q.~Qiao, J.~Cao, Y.~Wang and S.~Ye, \emph{Electrochimica Acta},
  2015, \textbf{176}, 1464--1475\relax
\mciteBstWouldAddEndPuncttrue
\mciteSetBstMidEndSepPunct{\mcitedefaultmidpunct}
{\mcitedefaultendpunct}{\mcitedefaultseppunct}\relax
\EndOfBibitem
\bibitem[Quintana and West(1987)]{QUINTANA1987179}
P.~Quintana and A.~West, \emph{Solid State Ionics}, 1987, \textbf{23},
  179--182\relax
\mciteBstWouldAddEndPuncttrue
\mciteSetBstMidEndSepPunct{\mcitedefaultmidpunct}
{\mcitedefaultendpunct}{\mcitedefaultseppunct}\relax
\EndOfBibitem
\bibitem[Mukai and Nunotani(2016)]{Mukai_2016}
K.~Mukai and N.~Nunotani, \emph{Journal of The Electrochemical Society}, 2016,
  \textbf{163}, A2371\relax
\mciteBstWouldAddEndPuncttrue
\mciteSetBstMidEndSepPunct{\mcitedefaultmidpunct}
{\mcitedefaultendpunct}{\mcitedefaultseppunct}\relax
\EndOfBibitem
\bibitem[QUI(1989)]{QUINTANA1989149}
\emph{Solid State Ionics}, 1989, \textbf{34}, 149--155\relax
\mciteBstWouldAddEndPuncttrue
\mciteSetBstMidEndSepPunct{\mcitedefaultmidpunct}
{\mcitedefaultendpunct}{\mcitedefaultseppunct}\relax
\EndOfBibitem
\bibitem[Wang \emph{et~al.}(2020)Wang, Aoyagi, Wisesa, and Mueller]{BOF}
C.~Wang, K.~Aoyagi, P.~Wisesa and T.~Mueller, \emph{Chemistry of Materials},
  2020, \textbf{32}, 3741--3752\relax
\mciteBstWouldAddEndPuncttrue
\mciteSetBstMidEndSepPunct{\mcitedefaultmidpunct}
{\mcitedefaultendpunct}{\mcitedefaultseppunct}\relax
\EndOfBibitem
\bibitem[Pilz and Jansen(2011)]{pilz}
T.~Pilz and M.~Jansen, \emph{Zeitschrift für anorganische und allgemeine
  Chemie}, 2011, \textbf{637}, 2148--2152\relax
\mciteBstWouldAddEndPuncttrue
\mciteSetBstMidEndSepPunct{\mcitedefaultmidpunct}
{\mcitedefaultendpunct}{\mcitedefaultseppunct}\relax
\EndOfBibitem
\bibitem[Court-Castagnet(1993)]{court-castagnet_1993}
R.~Court-Castagnet, \emph{Solid State Ionics}, 1993, \textbf{61},
  327–334\relax
\mciteBstWouldAddEndPuncttrue
\mciteSetBstMidEndSepPunct{\mcitedefaultmidpunct}
{\mcitedefaultendpunct}{\mcitedefaultseppunct}\relax
\EndOfBibitem
\bibitem[Tsujimura \emph{et~al.}(2022)Tsujimura, Ito, Yoshida, Higashiyama,
  Aihara, Machida, Park, and
  Im]{tsujimura_ito_yoshida_higashiyama_aihara_machida_park_im_2022}
T.~Tsujimura, S.~Ito, K.~Yoshida, Y.~Higashiyama, Y.~Aihara, N.~Machida,
  Y.~Park and D.~Im, \emph{Solid State Ionics}, 2022, \textbf{383},
  115970\relax
\mciteBstWouldAddEndPuncttrue
\mciteSetBstMidEndSepPunct{\mcitedefaultmidpunct}
{\mcitedefaultendpunct}{\mcitedefaultseppunct}\relax
\EndOfBibitem
\bibitem[Chun \emph{et~al.}(2021)Chun, Shim, and Yu]{chun_shim_yu_2021}
G.~H. Chun, J.~H. Shim and S.~Yu, \emph{ACS Applied Materials amp; Interfaces},
  2021, \textbf{14}, 1241–1248\relax
\mciteBstWouldAddEndPuncttrue
\mciteSetBstMidEndSepPunct{\mcitedefaultmidpunct}
{\mcitedefaultendpunct}{\mcitedefaultseppunct}\relax
\EndOfBibitem
\bibitem[Bogomolov \emph{et~al.}(2004)Bogomolov, Pantyukhina, and
  Surin]{AlSiO41}
M.~Bogomolov, M.~Pantyukhina and A.~Surin, \emph{Russian Journal of
  Electrochemistry}, 2004, \textbf{40}, 1029--1034\relax
\mciteBstWouldAddEndPuncttrue
\mciteSetBstMidEndSepPunct{\mcitedefaultmidpunct}
{\mcitedefaultendpunct}{\mcitedefaultseppunct}\relax
\EndOfBibitem
\bibitem[Takeuchi \emph{et~al.}(2016)Takeuchi, Yamashita, and
  Kuriyama]{AlSiO42}
Y.~Takeuchi, T.~Yamashita and K.~Kuriyama, \emph{J Solid State Electrochem},
  2016, \textbf{20}, 1885--1888\relax
\mciteBstWouldAddEndPuncttrue
\mciteSetBstMidEndSepPunct{\mcitedefaultmidpunct}
{\mcitedefaultendpunct}{\mcitedefaultseppunct}\relax
\EndOfBibitem
\bibitem[Patil \emph{et~al.}(2012)Patil, Hankare, Garadkar, and
  Sasikala]{PATIL201266}
R.~Patil, P.~Hankare, K.~Garadkar and R.~Sasikala, \emph{Journal of Alloys and
  Compounds}, 2012, \textbf{523}, 66--71\relax
\mciteBstWouldAddEndPuncttrue
\mciteSetBstMidEndSepPunct{\mcitedefaultmidpunct}
{\mcitedefaultendpunct}{\mcitedefaultseppunct}\relax
\EndOfBibitem
\bibitem[A.~Rais(2005)]{Crferr}
I.~A. A.-O. A.~Rais, A. M.~Gismelseed, \emph{physica status solidi (b)}, 2005,
  \textbf{242}, \relax
\mciteBstWouldAddEndPuncttrue
\mciteSetBstMidEndSepPunct{\mcitedefaultmidpunct}
{\mcitedefaultendpunct}{\mcitedefaultseppunct}\relax
\EndOfBibitem
\bibitem[Pan \emph{et~al.}(2018)Pan, Duan, Lin, Zong, Tong, Li, and
  Wang]{PAN2018270}
X.~Pan, X.~Duan, X.~Lin, F.~Zong, X.~Tong, Q.~Li and T.~Wang, \emph{Journal of
  Alloys and Compounds}, 2018, \textbf{732}, 270--279\relax
\mciteBstWouldAddEndPuncttrue
\mciteSetBstMidEndSepPunct{\mcitedefaultmidpunct}
{\mcitedefaultendpunct}{\mcitedefaultseppunct}\relax
\EndOfBibitem
\bibitem[Liu \emph{et~al.}(2019)Liu, Luo, Zhang, Hu, Yi, Wang, Zhang, Liu,
  Wang, Hao, Liu, and Guo]{CrFeO}
H.~Liu, S.-h. Luo, D.-x. Zhang, D.-b. Hu, T.-F. Yi, Z.-y. Wang, Y.-h. Zhang,
  Y.-g. Liu, Q.~Wang, A.-m. Hao, X.-w. Liu and R.~Guo, \emph{ChemElectroChem},
  2019, \textbf{6}, 856--864\relax
\mciteBstWouldAddEndPuncttrue
\mciteSetBstMidEndSepPunct{\mcitedefaultmidpunct}
{\mcitedefaultendpunct}{\mcitedefaultseppunct}\relax
\EndOfBibitem
\bibitem[Sadan and Kaplan(2006)]{AuSaphh}
H.~Sadan and W.~Kaplan, \emph{J Mater Sci}, 2006, \textbf{41}, \relax
\mciteBstWouldAddEndPuncttrue
\mciteSetBstMidEndSepPunct{\mcitedefaultmidpunct}
{\mcitedefaultendpunct}{\mcitedefaultseppunct}\relax
\EndOfBibitem
\bibitem[Tran \emph{et~al.}(2016)Tran, Xu, Radhakrishnan, Winston, Sun,
  Persson, and Ong]{TRAN2016}
R.~Tran, Z.~Xu, B.~Radhakrishnan, D.~Winston, W.~Sun, K.~A. Persson and S.~P.
  Ong, \emph{Scientific Data}, 2016, \textbf{3}, \relax
\mciteBstWouldAddEndPuncttrue
\mciteSetBstMidEndSepPunct{\mcitedefaultmidpunct}
{\mcitedefaultendpunct}{\mcitedefaultseppunct}\relax
\EndOfBibitem
\bibitem[Tran \emph{et~al.}(2019)Tran, Li, Montoya, Winston, Persson, and
  Ong]{TRAN201948}
R.~Tran, X.-G. Li, J.~H. Montoya, D.~Winston, K.~A. Persson and S.~P. Ong,
  \emph{Surface Science}, 2019, \textbf{687}, 48--55\relax
\mciteBstWouldAddEndPuncttrue
\mciteSetBstMidEndSepPunct{\mcitedefaultmidpunct}
{\mcitedefaultendpunct}{\mcitedefaultseppunct}\relax
\EndOfBibitem
\bibitem[Schindler \emph{et~al.}(2020)Schindler, Antoniuk, Cheon, Zhu, and
  Reed]{peter}
P.~Schindler, E.~R. Antoniuk, G.~Cheon, Y.~Zhu and E.~J. Reed, \emph{arXiv},
  2020\relax
\mciteBstWouldAddEndPuncttrue
\mciteSetBstMidEndSepPunct{\mcitedefaultmidpunct}
{\mcitedefaultendpunct}{\mcitedefaultseppunct}\relax
\EndOfBibitem
\bibitem[Hatzell and Zheng(2021)]{hatzell_zheng_2021}
K.~B. Hatzell and Y.~Zheng, \emph{MRS Energy amp; Sustainability}, 2021,
  \textbf{8}, 33–39\relax
\mciteBstWouldAddEndPuncttrue
\mciteSetBstMidEndSepPunct{\mcitedefaultmidpunct}
{\mcitedefaultendpunct}{\mcitedefaultseppunct}\relax
\EndOfBibitem
\end{mcitethebibliography}
\bibliographystyle{rsc} 

\end{document}